\documentclass[preprint2]{aastex}

\usepackage{calrsfs,euscript,mathrsfs,amssymb}
\usepackage{graphicx}
\usepackage[nomarkers,nolists]{endfloat}
\usepackage{lineno}
\newcount\Comments  % 0 suppresses notes to selves in text
\usepackage[usenames,dvipsnames]{xcolor}
\definecolor{darkgreen}{rgb}{0,0.5,0}
\definecolor{purple}{rgb}{1,0,1}
\definecolor{darkpurple}{rgb}{0.5,0,0.5}
\definecolor{lightgreen}{RGB}{135,220,0}

\newcommand{\kibitz}[2]{\ifnum\Comments=1\textcolor{#1}{#2}\fi}
\newcommand{\lat}{\textit{Fermi}-LAT }
\newcommand{\blat}{$|b|<10{^\circ}$ }
\newcommand{\blot}{$|b|\geq 10{^\circ}$ }
\newcommand{\eg}{e.g.~}
\newcommand{\ie}{i.e.~}

\shorttitle{2FHL Catalog}
\shortauthors{Ajello, Dom{\'i}nguez, }
\slugcomment{}

\begin{document}


\title{2FHL: The Second Catalog of Hard \lat  Sources}
%\email{Version 6.0}
%\title{The Second Fermi-LAT Catalog of Hard Sources\\Author list created Thursday 30 Jul 2015 01:18 PDT}
\author{
M.~Ackermann\altaffilmark{2}, 
M.~Ajello\altaffilmark{3,1}, 
W.~B.~Atwood\altaffilmark{4}, 
L.~Baldini\altaffilmark{5,6}, 
J.~Ballet\altaffilmark{7}, 
G.~Barbiellini\altaffilmark{8,9}, 
D.~Bastieri\altaffilmark{10,11}, 
R.~Bellazzini\altaffilmark{12}, 
E.~Bissaldi\altaffilmark{13}, 
R.~D.~Blandford\altaffilmark{6}
% ...
}
\altaffiltext{1}{Corresponding authors: M.~Ajello, majello@clemson.edu;
 A.~Dom\'inguez, alberto@clemson.edu;
 J.~Cohen, jcohen@astro.umd.edu; 
S.~Cutini, sara.cutini@asdc.asi.it; 
D.~Gasparrini, gasparrini@asdc.asi.it.}
\altaffiltext{2}{Deutsches Elektronen Synchrotron DESY, D-15738 Zeuthen, Germany}
\altaffiltext{3}{Department of Physics and Astronomy, Clemson University, Kinard Lab of Physics, Clemson, SC 29634-0978, USA}
\altaffiltext{4}{Santa Cruz Institute for Particle Physics, Department of Physics and Department of Astronomy and Astrophysics, University of California at Santa Cruz, Santa Cruz, CA 95064, USA}
\altaffiltext{5}{Universit\`a di Pisa and Istituto Nazionale di Fisica Nucleare, Sezione di Pisa I-56127 Pisa, Italy}
\altaffiltext{6}{W. W. Hansen Experimental Physics Laboratory, Kavli Institute for Particle Astrophysics and Cosmology, Department of Physics and SLAC National Accelerator Laboratory, Stanford University, Stanford, CA 94305, USA}
\altaffiltext{7}{Laboratoire AIM, CEA-IRFU/CNRS/Universit\'e Paris Diderot, Service d'Astrophysique, CEA Saclay, F-91191 Gif sur Yvette, France}
\altaffiltext{8}{Istituto Nazionale di Fisica Nucleare, Sezione di Trieste, I-34127 Trieste, Italy}
\altaffiltext{9}{Dipartimento di Fisica, Universit\`a di Trieste, I-34127 Trieste, Italy}
\altaffiltext{10}{Istituto Nazionale di Fisica Nucleare, Sezione di Padova, I-35131 Padova, Italy}
\altaffiltext{11}{Dipartimento di Fisica e Astronomia ``G. Galilei'', Universit\`a di Padova, I-35131 Padova, Italy}
\altaffiltext{12}{Istituto Nazionale di Fisica Nucleare, Sezione di Pisa, I-56127 Pisa, Italy}
\altaffiltext{13}{Istituto Nazionale di Fisica Nucleare, Sezione di Bari, I-70126 Bari, Italy}
% ...

\begin{abstract}
We present a catalog of sources detected above  50\,GeV by the {\it Fermi}-Large Area Telescope (LAT) in 80\,months of data. The newly delivered Pass~8 event-level analysis  allows the detection and characterization of sources in the 50\,GeV--2\,TeV energy range. In this energy band, \lat has detected 360 sources, { which} constitute the second catalog of hard \lat sources (2FHL). The improved angular resolution { enables} the precise localization of point sources ($\sim$1.7$'$ radius at 68\%\,C.~L.) and the detection and characterization of spatially extended sources. We find that 86\,\% of the sources can be associated with counterparts at other wavelengths, of which the majority (75\,\%) are active galactic nuclei and the rest (11\,\%) are Galactic sources. Only 25\,\% of the 2FHL sources { have} been previously detected by Cherenkov telescopes, implying that the 2FHL provides a reservoir of candidates to be followed up at very high energies. This work closes the energy gap between the observations performed at GeV energies by \lat on orbit and the observations performed at higher energies by Cherenkov telescopes from the ground.

\end{abstract}

\keywords{catalogs -- gamma rays: general}

%%%%%%%%%%%%%%%%%%%%%%%%%%%%%%%%%%%%%%%%%%%%%%%%%%%%%%%%%%%%%%%%
%
%         Introduction 
%
%%%%%%%%%%%%%%%%%%%%%%%%%%%%%%%%%%%%%%%%%%%%%%%%%%%%%%%%%%%%%%%%
\section{Introduction}

The Large Area Telescope \citep[LAT,][]{atwood09} on board the {\it Fermi}
$\gamma$-ray space telescope has been surveying the whole sky since August 2008.
Its unprecedented sensitivity and localization accuracy allowed the detection
of over 3,000 point-like sources in  4\,years of data \citep[see the
third catalog of {\it Fermi}-LAT sources, 3FGL, ][]{3FGL}.
Typically, {\it Fermi}-LAT catalog studies are based on  source detection
and characterization in the whole 0.1\,GeV--100\,GeV energy band.
The larger photon statistics present at low energy, counterbalanced by
the LAT point-spread function (PSF) whose size decreases with energy,
yields an optimum sensitivity at few-GeV energies.
The {\it Fermi}-LAT catalogs are thus representative of the GeV sky
more than they are of the MeV or the sub-TeV sky.

The first {\it Fermi}-LAT catalog of hard sources, named 1FHL \citep{1FHL}, provided an unbiased census of the sky at energies from 10~GeV up to 500~GeV. The comparison of 1FHL and 0.1--100\,GeV observations \citep[as provided in][]{2FGL} allowed us to uncover the presence of spectral breaks and to determine
that  blazars of the BL Lacertae (BL Lac) type represented about 50\,\%
of the entire source population { detected in that band}.
{ All-sky surveys at $\gamma$-ray energies} are instrumental for ground-based imaging atmospheric Cherenkov telescopes (IACTs) such as H.E.S.S., MAGIC, and VERITAS \citep[][respectively]{holder08,lorenz04,hinton04}
in order to find new sources because of their limited fields of view (FoV).

Recently, a new event-level  analysis (known as Pass 8) has been developed by the \emph{Fermi}-LAT collaboration \citep{atwood13b,atwood13}. Pass~8 significantly improves the LAT's background rejection, PSF, and effective area. All these  enhancements lead to a significant increase of the LAT sensitivity and its effective energy range, from below 100~MeV to beyond a few hundred GeV  \citep{atwood13b,atwood13}. These improvements are particularly significant above 50\,GeV, yielding an enhancement in the acceptance and PSF by a factor  between 1.2 and 2. 
It is interesting to note that, above 50\,GeV, both the PSF (governed mostly by the pitch of the tracker silicon strips and the spacing of the tracker planes) and the effective area of the LAT are only weakly dependent on energy and that the LAT 
 operates, due to the (almost complete) absence of background, in the photon-limited regime.

In this paper we use 80\,months of Pass~8 data to produce a catalog of sources detected by the LAT at energies\footnote{Note the different energy range with respect to the 1FHL.} between 50\,GeV and 2\,TeV. This constitutes the second catalog of hard \lat sources, named 2FHL, which allows a thorough study of the properties of the whole sky in the sub-TeV domain.

The paper is organized as follows: $\S$\ref{sec:analysis} describes the analysis,
and $\S$\ref{sec:results} discusses the 2FHL catalog and the main results.
A summary is provided in $\S$\ref{sec:summary}.

%%%%%%%%%%%%%%%%%%%%%%%%%%%%%%%%%%%%%%%%%%%%%%%%%%%%%%%%%%%%%%%%
%
%         Analysis 
%
%%%%%%%%%%%%%%%%%%%%%%%%%%%%%%%%%%%%%%%%%%%%%%%%%%%%%%%%%%%%%%%%
\section{Analysis}
\label{sec:analysis}

%%%%%%%%%%%%%%%%%%%%%%%%%%%%%%%%%%%%%%%%%%%%%%%%%%%%%%%%%%%%%%%%
%
%         Detection
%
%%%%%%%%%%%%%%%%%%%%%%%%%%%%%%%%%%%%%%%%%%%%%%%%%%%%%%%%%%%%%%%%

\subsection{\label{sec:data_sel}Data Selection}

\begin{figure*}[!ht]
\begin{centering} 
\includegraphics[scale=0.85]{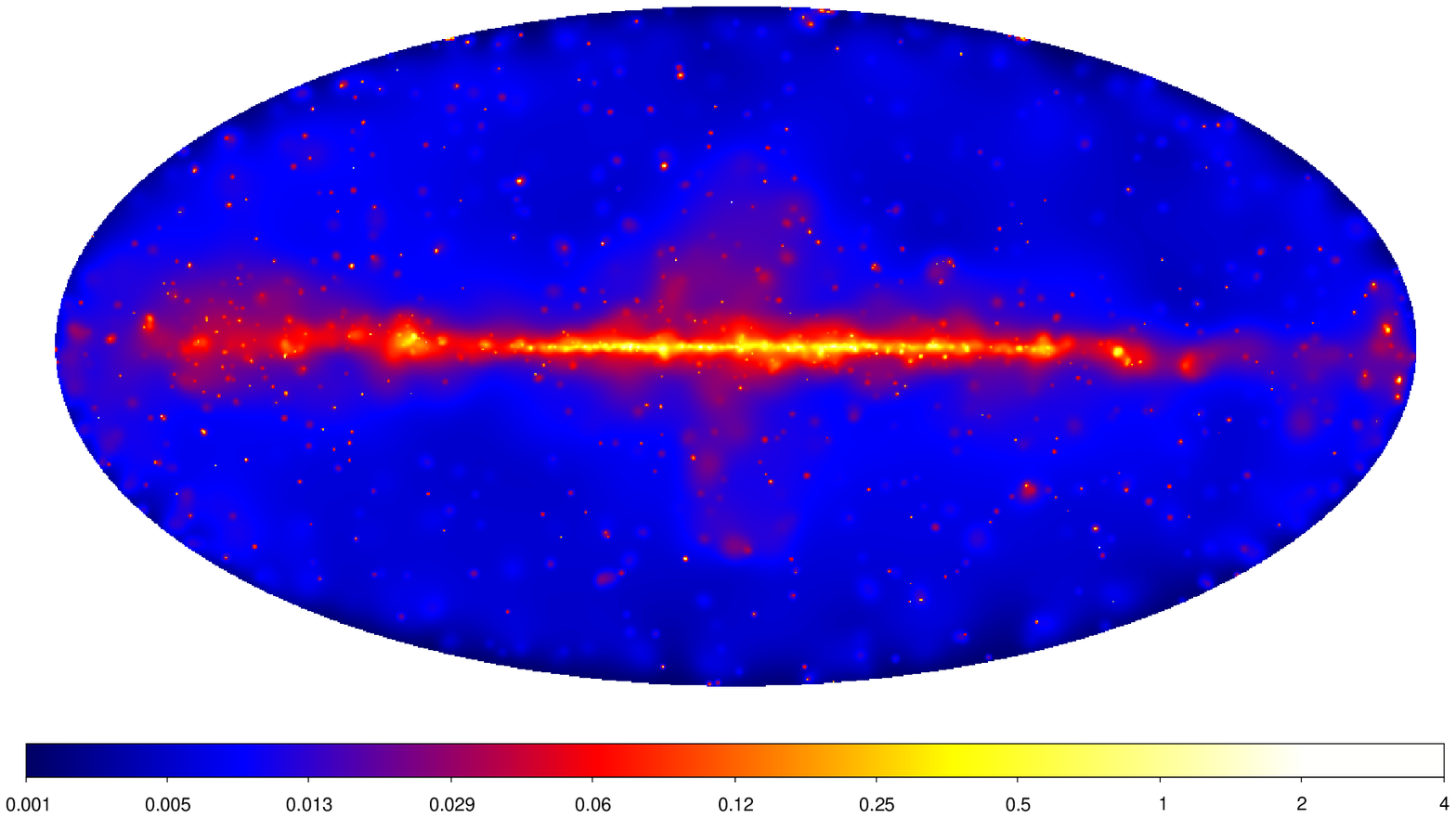}
\caption{Adaptively smoothed count map in the 50\,GeV--2\,TeV band represented in Galactic coordinates and Hammer-Aitoff projection. The image has been smoothed with a Gaussian kernel whose size was varied to achieve a minimum signal-to-noise ratio under the kernel of 2. The color scale is logarithmic and the units are counts per (0.1\,deg)$^2$.
    \label{fig:skymap}}
\end{centering}
\end{figure*}

We use 80 months (from August 2008 to April 2015) of P8\_SOURCE 
photons with reconstructed energy in the 50\,GeV--2\,TeV range.
At these energies the LAT has an energy resolution of around 10--15\,\% (1\,$\sigma$).
Photons detected at zenith angles larger than 105$^{\circ}$ were excised
to limit the contamination from $\gamma$-rays generated by cosmic-ray
interactions in the upper layers of the atmosphere. Moreover, data were filtered
removing time periods when the instrument was not in sky-survey mode\footnote{This
was achieved using the expression `(DATA\_QUAL$>$0)\&\&(LAT\_CONFIG==1)' in {\tt gtmktime}.}.
This leaves
approximately 61,000 photons detected all over the sky. The count map
reported in Figure~\ref{fig:skymap} shows that \lat observes
many point-like sources and
 large scale diffuse emission in the direction of our Galaxy, some of which appears
coincident  with the so-called {\it Fermi} bubbles \citep{su10,lat_bubbles}.

\subsection{\label{sec:sourceDetect}Source Detection}

The first step of the source detection stage comprises the identification of source
seeds, which are locations of potential sources whose significance is
later tested
through a maximum likelihood (ML) analysis. Seeds are identified 
via a sliding-cell algorithm as excesses above the background, 
as clusters of 3 or more photons, and via a wavelet analysis
\citep{ciprini07}. Moreover, the seed list
includes all the point sources detected in the 1FHL catalog.
We note that  this seed list may include statistical fluctuations as well as real sources with a non-optimal position.

A full ML analysis is then performed in order to verify which,
 among the seeds, are the reliable sources. 
The analysis is performed in 154 regions of interest (ROIs), varying between 10$^{\circ}$ and 20$^{\circ}$ in radius, whose sizes and positions in the sky are optimized to cover all the seeds, ensuring that no more than 45 seeds are contained in a single ROI.
 For each ROI, we build
a sky model that includes all the potential sources in the region
as well as the  Galactic and isotropic diffuse emissions\footnote{We~use~the~{\tt gll\_iem\_v06.fits}~and~{\tt iso\_P8R2\_SOURCE\_V6\_v06.txt}
templates available at \\
http://fermi.gsfc.nasa.gov/ssc/data/access/lat/BackgroundModels.html.}.
These models, which are defined only up to $\sim$600\,GeV and  $\sim$900\,GeV respectively, where extrapolated up to 2\,TeV.
The ROI models include also the extended sources present in the region (see $\S$~\ref{sec:extended}).
The model is fit to the data via the unbinned ML algorithm provided
within the {\it Fermi} Science Tools\footnote{Available at 
http://fermi.gsfc.nasa.gov/ssc/data/analysis/software/.} (version v9r34p3).

The spectrum of each source is modeled with a power law because
none of the sources is expected to show statistically
significant spectral
curvature detectable by the LAT in this energy band. 
Indeed, this was the case for the sources in the 1FHL catalog \citep{1FHL}.

The fit is performed iteratively in order to ensure convergence and 
to produce an optimal solution. It proceeds as follows:

\begin{enumerate}
\item Complex ML fits require approximate knowledge of the 
starting values of the parameters. For this reason the first
step  aims to find those values  by fitting each single source separately 
 to determine approximate spectral parameters.
Throughout the entire process, the parameters of the diffuse emission
models are left free to vary.
The significance of each source is evaluated using the test statistic  
${\rm TS}=2(\ln \mathcal{L}_1 - \ln \mathcal{L}_0)$, where $\mathcal{L}_0$ and $\mathcal{L}_1$
are the likelihoods of the background (null hypothesis) and 
the hypothesis being tested (\eg source plus background). 
At each step in the procedure, marginal sources, those with TS$<$10, are removed from the model.
Once the spectral parameters and significance of each source have
been evaluated, a global fit for which all the parameters of the sources
with a ${\rm TS}\geq 10$ are allowed to  vary is performed. 
Then one more global fit is performed after removing all the sources that had ${\rm TS}< 10$ at the previous global fit.
This step,
as well as all the others, includes sources that
are spatially extended (see $\S$\ref{sec:extended});

\item In this second step, the positions of point-like sources, using the best-fit sky model
derived at step 1, are optimized using the {\tt gtfindsrc} tool. This step
is done iteratively as well by optimizing first the positions of the
most significant sources found at step 1 and later  those of the fainter
ones;

\item The parameters and significances of sources are estimated again (as in step 1) using the best-fit source positions. This step produces the best-fit sky 
model for any given ROI. 
Seeds with $10\leq {\rm TS}<25$ are included in the model, but not reported in the final catalog;

\item For each source we estimate the energy of the highest-energy photon (HEP)
that the fit attributes robustly  to the source model. This is done using the tool {\tt gtsrcprob} and selecting the HEP that has a probability $>85$\,\% to belong to the source;

\item A  spectrum with three logarithmically spaced bins (boundaries of 50\,GeV, 171\,GeV, 585\,GeV, 2\,TeV) is generated for each source in the ROI that is detected with ${\rm TS}\geq 25$ and with the number of detected $\gamma$ rays (estimated by the likelihood, N$_{\rm pred}$) to be $\geq$3.

\end{enumerate}

The procedure described above achieves the detection of 360 sources (including
the extended sources discussed next at $\S$~\ref{sec:extended}) 
with TS $\geq$ 25 and N$_{\rm pred}\geq3$ across the entire sky.
The number of seeds kept in the ROI models
with $10\leq{\rm TS}<25$ is 453, while 7 are the seeds with ${\rm TS\geq}25$, but N$_{\rm pred}<3$.
 We have performed seven Monte Carlo
simulations of the $>$50\,GeV sky whose data have been analyzed 
like the real data (as detailed above). The  N$_{\rm pred}$ cut was introduced
on the basis of these simulations to limit to $\lesssim$1\,\% the number of false positives in the final catalog. These simulations will be discussed
in a forthcoming publication.

%%%%%%%%%%%%%%%%%%%%%%%%%%%%%%%%%%%%%%%%%%%%%%%%%%%%%%%%%%%%%%%%
%
%         Extended Sources
%
%%%%%%%%%%%%%%%%%%%%%%%%%%%%%%%%%%%%%%%%%%%%%%%%%%%%%%%%%%%%%%%%
\subsection{Search for Spatially-Extended Sources}
\label{sec:extended}

Preliminary runs of the source detection method described in $\S$~\ref{sec:sourceDetect} detected clusters of point sources in the Galactic plane, which were suggestive of spatially extended sources. It is also possible that clusters of seed sources, each with sub-detection-threshold significance, could be detected as a significant extended source. 
 Not modeling extended $\gamma$-ray emission as such can lead to inaccurate measurements of spectral and spatial properties of both the extended source and neighboring point sources, particularly in the Galactic plane \citep{Lande12}. Most of the TeV sources in the Galactic plane are spatially
extended  \citep{carrigan2013, ong2013}, 
so to clearly connect LAT detections spectrally to these sources, extension detection and characterization is important.
In the following, we distinguish between sources whose extension
has been previously determined with {\it Fermi}-LAT and
new extended sources that are reported for the first time
in a \lat catalog. The details of all significantly
detected extended sources will be reported  in 
$\S~$\ref{sec:ESresults}. 

%%%%%%%%%%%%%%%%%%%%%%%%%%%%%%%%%%%%%%%%%%%%%%%%%%%%%%%%%%%%%%%%
%
%         Extended Sources in 3FGL
%
%%%%%%%%%%%%%%%%%%%%%%%%%%%%%%%%%%%%%%%%%%%%%%%%%%%%%%%%%%%%%%%%

\subsubsection{\label{sec:3FGL_ES}Extended Sources Previously Detected by the LAT}
We explicitly modeled sources as spatially extended when a previous, dedicated, analysis found the source to be resolved by the LAT.
The 25 extended sources reported in 3FGL were included in our model using the spatial templates derived in the individual source studies \citep[see references in ][]{3FGL}. Refitting the positions and extensions of the 3FGL extended sources in this energy range is beyond the scope of this work.

Of the 25 3FGL extended sources, 19 are significantly detected here above the detection threshold (${\rm TS}\geq25$). Only 6 sources are not detected and, since all have  ${\rm TS<10}$, are removed from the sky model (see \S\ref{sec:ESresults} for details).

One extended LAT source has had a dedicated analysis published since the release of the 3FGL catalog. \cite{HESSLATW41} reported joint H.E.S.S. and LAT observations of the very high energy (VHE) source HESS~J1834-087. This source is coincident with supernova remnant (SNR) W41 and was detected  as spatially extended in a wide energy range spanning 1.8\,GeV to 30\,TeV. In this paper, we employ the spatial model for the GeV emission determined in \cite{HESSLATW41}, leading to a significant detection of this source.

%%%%%%%%%%%%%%%%%%%%%%%%%%%%%%%%%%%%%%%%%%%%%%%%%%%%%%%%%%%%%%%%
%
%         Extended Sources not in 3FGL
%
%%%%%%%%%%%%%%%%%%%%%%%%%%%%%%%%%%%%%%%%%%%%%%%%%%%%%%%%%%%%%%%%
\subsubsection{\label{sec:newES}Newly Detected Extended Sources} 

In addition to modeling the extended sources mentioned in $\S$\ref{sec:3FGL_ES}, we performed a blind search of the Galactic plane  (\blat) to identify potential extended sources not included in previously published works. Our analysis pipeline is similar to that used in \cite{hewittSNRcat13}, with some modifications tailored to searching for multiple extended sources in an ROI. The pipeline employs the {\tt pointlike} binned maximum likelihood package \citep{Kerr10}, in particular utilizing the extended source fitting tools validated by \cite{Lande12} to simultaneously fit the position, extension, and spectra of sources in our ROI. 

We created 72 ROIs of radius ${\rm 10^{\circ}}$, centered on ${\rm b = 0^{\circ}}$ with neighboring ROIs overlapping and separated by ${\rm 5^{\circ}}$ in Galactic longitude.  Our initial model of the $\gamma$-ray emission in each ROI consisted solely of the Galactic diffuse (allowing just the normalization to be fit) and isotropic emission models (fixing the normalization), with no other sources in the ROI. Emission in the ROIs was further characterized by adding sources and fitting their spectral parameters (normalization and spectral index) in a ${\rm 14^{\circ}}\times{\rm 14^{\circ}}$ region. 

A TS map, that included all significant sources found previously, made up of ${\rm 0.1^{\circ}}\times{\rm 0.1^{\circ}}$ bins across the ROI, was created at each iteration and a small radius (${\rm 0.1^{\circ}}$) uniform disk, with a power-law spectrum was placed at the position of the peak TS pixel. The spectra of any newly added sources, as well as the position, extension, and spectral parameters of the disk were then fit. If $\rm TS_{ext} \geq 16$, where  ${\rm TS_{ext} = 2~log(\mathcal{L}_{ext} / \mathcal{L}_{ps})}$ \citep[\ie twice the logarithm likelihood ratio of an extended to a point source,][]{Lande12}, then the disk was kept in the model. For $\rm TS_{ext} < 16$, the extended source was replaced by a point source with a power-law spectral model. For the point-source replacement case, spectral parameters of sources in the ROI were fit and the position of the new point source was optimized. Finally, the spatial parameters of any previously added extended sources were refit iteratively before creating a new TS map and repeating the process. We stopped adding sources when the peak TS was less than 16 for two successive sources. 

{ To assess the impact of fitting extended sources when starting with an ROI devoid of sources, a crosscheck analysis (also using {\tt pointlike}) was performed across the Galactic plane. We included 3FGL point and extended sources, the Galactic diffuse and isotropic emission, and pulsars from the second \lat pulsar catalog \citep{2PC} (as well as from 3FGL) in the preliminary source model for each region. Sources were iteratively added to account for residual emission and both these residual sources and 3FGL  sources were tested for extension. Remarkably, this alternative analysis converges (\ie spectral and spatial parameters for the detected extended sources are compatible in both analyses) to the initially source-devoid analysis for nearly all detected extended sources.
}

Extended sources detected in the analysis described in this section for which the position and extension  were compatible with those found by the crosscheck were included in the ROI model at step 1 of the full ML analysis detailed in \S\ref{sec:sourceDetect}. Seed point sources interior to the extended sources were removed prior to the ML fit. { To address the ambiguity between detecting a source as spatially extended as opposed to a combination of point sources, we utilized the algorithm detailed in \cite{Lande12} to simultaneously fit the spectra and positions of two nearby point sources. We only consider a source to be extended if $\rm TS_{ext}~ >$ $\rm TS_{2pts}$  (improvement when adding a second point source).} Our blind search of the Galactic plane allowed us to find 5 sources not previously detected as extended by {\it Fermi}-LAT.  Further details on these sources are presented in \S~\ref{sec:ESresults}.

%%%%%%%%%%%%%%%%%%%%%%%%%%%%%%%%%%%%%%%%%%%%%%%%%%%%%%%%%%%%%%%%
%
%         Comparison with Pass 7
%
%%%%%%%%%%%%%%%%%%%%%%%%%%%%%%%%%%%%%%%%%%%%%%%%%%%%%%%%%%%%%%%%
\subsection{Comparison with Pass 7}

In order to gauge the improvement delivered by Pass~8 at $\geq$50\,GeV,
we repeated the analysis procedure described above with
 80 months of Pass~7 reprocessed data. This analysis detected $\sim$230 sources; $\sim$35\,\% fewer
than the corresponding analysis that relies on Pass~8 data.
The main difference is for the region $|b|>10^{\circ}$ where Pass~8 data, because of the larger acceptance, and  better PSF  
allow the detection of $\sim$60\,\% more sources
than what could be achieved with Pass~7. 
The improvements delivered
by Pass~8 above 50\,GeV are clearly substantial.

%%%%%%%%%%%%%%%%%%%%%%%%%%%%%%%%%%%%%%%%%%%%%%%%%%%%%%%%%%%%%%%%
%
%         Associations
%
%%%%%%%%%%%%%%%%%%%%%%%%%%%%%%%%%%%%%%%%%%%%%%%%%%%%%%%%%%%%%%%%
\subsection{Source Association}
\label{sec:associations}

The approach for automated source association closely
follows that used for the 2FGL, 1FHL,  3FGL and 3LAC catalogs \citep{2FGL,1FHL,3FGL,3LAC}. In short, we use catalogs  of known or potential $\gamma$-ray source classes to determine the probability that
a source from a given catalog or survey
is associated with a 2FHL source.

The associations were derived with two different procedures:
the Bayesian method and the likelihood-ratio  method
\citep[described in detail in ][]{3LAC,3FGL}.
 In the application of these two methods, potential counterparts
were deemed  associated if they were found to have a posteriori probability
of at least 80\,\%.

For the Bayesian method, the catalogs relavant for associating 2FHL sources are the 5th version of the BZCAT \citep{bzcat5} and the ATNF Pulsar Catalog \citep{Manchester2005}. Other catalogs of Galactic populations (X-ray binaries, O stars, Wolf-Rayet stars, luminous blue variable stars, open and globular clusters) were used in the procedure, but no counterparts reached the probability threshold.

For the  likelihood-ratio association method, we made use of a number of relatively uniform radio surveys.  Almost all radio candidates of possible interest are in the  NRAO VLA Sky Survey  \citep[NVSS; ][]{condon98} or the Sydney University Molonglo Sky Survey \citep[SUMSS; ][]{mauch03}.  To look for additional possible counterparts we cross-correlated the LAT sources with the most sensitive all-sky X-ray survey, the ROSAT All Sky Survey Bright and Faint Source Catalogs \citep{voges99,voges00}. 

In order to be consistent, we evaluated matches between the 2FHL and the 3FGL and 1FHL catalogs, cross correlating those catalogs with the 2FHL one taking
into account the source location uncertainty (95\,\% confidence) regions.
In those cases, we adopted the source associations given in  previously
published \lat catalogs. { This yielded only two associations (Eta Carinae and 
the SNR G338.3-0.0), which were both detected and associated in both the 3FGL and 1FHL.}
The 2FHL catalog was also cross-correlated with the TeVCat\footnote{See http://tevcat.uchicago.edu/ .}(which contains all the TeV sources detected by IACTs so far)  and spatial coincidences are reported (see $\S$~\ref{sec:tevcat}). 
Since they have no positional error associated, the association probability was not computed for the extended sources.

High-confidence associations 
allow us to assess the systematic uncertainty in the accuracy of the LAT source
positions.
As done in \cite{3FGL}, we compared the distribution of angular
separations of the $\gamma$-ray sources to the highest-confidence (probability $>$90\,\%)
counterparts (in units of 1\,$\sigma$ errors) with a Rayleigh distribution,
and found it slightly broader than expected. Consequently, the standard 68\,\% uncertainty
radius provided by {\tt gtfindsrc} has been multiplied by { 1.08 and summed in quadrature to 0$^{\circ}$.003 (68\% absolute systematic error from 3FGL).  }

% Table listing the source classes and their numbers
\begin{deluxetable}{lcr}
\setlength{\tabcolsep}{0.04in}
\tablewidth{0pt}
\tabletypesize{\small}
\tablecaption{2FHL Source Classes \label{tab:class}}
\tablehead{
\colhead{Description} & 
\multicolumn{2}{c}{Associated} \\
& 
\colhead{Designator} &
\colhead{Number}
}
\startdata
Pulsar & psr & 1 \\
Pulsar wind nebula & pwn & 14 \\
Supernova remnant & snr & 16 \\
Supernova remnant / Pulsar wind nebula & spp & 4 \\
High-mass binary & hmb & 2 \\
Binary & bin & 1 \\
Star-forming region & sfr & 1 \\
BL Lac type of blazar & bll & 180 \\
BL Lac type of blazar with prominent galaxy emission & bll-g & 13 \\
FSRQ type of blazar & fsrq & 10 \\
Non-blazar active galaxy & agn & 2 \\ 
Radio galaxy & rdg & 4 \\
Radio galaxy / BL Lac  & rdg/bll & 2 \\
Blazar candidate of uncertain type I & bcu I & 7 \\
Blazar candidate of uncertain type II & bcu II & 34 \\ 
Blazar candidate of uncertain type III & bcu III & 19 \\  
Normal galaxy (or part) & gal & 1 \\
Galaxy cluster & galclu & 1 \\
Total associated & \nodata & 312 \\
%\hline
Unassociated & \nodata & 48 \\ 
\hline
Total in 2FHL & \nodata & 360 \\ 

\enddata
\tablecomments{The designation `spp' indicates potential association with SNR or PWN.
The `bcu I', `bcu II', and `bcu III' classes are derived from 3LAC  and describe the increasing lack of multiwavelength information to classify  the source as a blazar \citep[see ][for more details]{3LAC}. The designation `bll-g' is adapted from the BZCAT \citep{bzcat5} and indicates a blazar whose SED has a significant contribution from the host galaxy.}
\end{deluxetable}
 % TAB 1

\begin{deluxetable}{lccl}
\setlength{\tabcolsep}{0.04in}
\tablewidth{0pt}
\tabletypesize{\scriptsize}
\tablecaption{Description of the catalog \label{tab:description}}
\tablehead{
\colhead{Column} & 
\colhead{Format} &
\colhead{Unit} &
\colhead{Description}
}
\startdata
Source\_Name & 18A & \nodata & 2FHL Source name \\
RAJ2000 & E & deg & Right ascension \\
DEJ2000 & E & deg & Declination \\
GLON & E & deg & Galactic longitude \\
GLAT & E & deg & Galactic latitude \\
Pos\_Err\_68 & E & deg & Position uncertainty at 68\% confidence level \\
TS & E & \nodata & Test Statistic\\
Spectral\_Index & E & \nodata & Observed spectral index \\
Unc\_Spectral\_Index & E & \nodata & 1$\sigma$ uncertainty on the observed spectral index \\
Flux50 & E & photon~cm$^{-2}$~s$^{-1}$ & Integral photon flux from 50~GeV to 2~TeV \\
Unc\_Flux50 & E & photon~cm$^{-2}$~s$^{-1}$ & 1$\sigma$ uncertainty on integral photon flux from 50~GeV to 2~TeV \\
Energy\_Flux50 & E & erg~cm$^{-2}$~s$^{-1}$ & Energy flux from 50~GeV to 2~TeV  \\
Unc\_Energy\_Flux50 & E & erg~cm$^{-2}$~s$^{-1}$ & 1$\sigma$ uncertainty on energy flux from 50~GeV to 2~TeV  \\
Flux50\_171GeV & E & photon~cm$^{-2}$~s$^{-1}$ & Integral photon flux from 50~GeV to 171~GeV \\
Unc\_Flux50\_171GeV & E & photon~cm$^{-2}$~s$^{-1}$ & 1$\sigma$ uncertainty on integral photon flux from 50~GeV to 171~GeV \\
Sqrt\_TS50\_171GeV & E & photon~cm$^{-2}$~s$^{-1}$ & Square root of the Test Statistic between 50~GeV and 171~GeV \\
Flux171\_585GeV & E & photon~cm$^{-2}$~s$^{-1}$ & Integral photon flux from 171~GeV to 585~GeV \\
Unc\_Flux171\_585GeV & E & photon~cm$^{-2}$~s$^{-1}$ & 1$\sigma$ uncertainty on integral photon flux from 171~GeV to 585~GeV \\
Sqrt\_TS171\_585GeV & E & photon~cm$^{-2}$~s$^{-1}$ & Square root of the Test Statistic between 171~GeV and 585~GeV \\
Flux585\_2000GeV & E & photon~cm$^{-2}$~s$^{-1}$ & Integral photon flux from 585~GeV to 2~TeV \\
Unc\_Flux585\_2000GeV & E & photon~cm$^{-2}$~s$^{-1}$ & 1$\sigma$ uncertainty on integral photon flux from 585~GeV to 2~TeV \\
Sqrt\_TS585\_2000GeV & E & photon~cm$^{-2}$~s$^{-1}$ & Square root of the Test Statistic between 585~GeV and 2~TeV \\
Npred & E & \nodata & Predicted number of photons from the source \\
HEP\_Energy & E & GeV & Highest photon energy \\
HEP\_Prob & E & \nodata & Probability that the highest-energy photon is coming from the source, $\geq 0.85$ \\
ROI & E & \nodata & Region of interest number \\
ASSOC & 25A & \nodata & Name of the most likely associated source \\
ASSOC\_PROB\_BAY & E & \nodata & Probability of association from the Bayesian method\\
ASSOC\_PROB\_LR & E  & \nodata & Probability of association from the likelihood ratio method\\
CLASS & 8A & \nodata & Class designation for the most likely association; see Table~\ref{tab:class} \\
Redshift & E & \nodata & Redshift (when available) of the most likely associated source \\
NuPeak\_obs & E & Hz & Observed Synchrotron peak frequency \\
3FGL\_Name & 18A & \nodata & Name of the most likely associated source in the 3FGL\\
1FHL\_Name & 18A & \nodata & Name of the most likely associated source in the 1FHL \\
TeVCat\_Name & 18A & \nodata & Name of the most likely associated source in the TeVCat\\
\enddata
\tablecomments{A 'Source\_Name' ending with 'e' indicates an extended source.}
\end{deluxetable}
 % TAB 2

%%%%%%%%%%%%%%%%%%%%%%%%%%%%%%%%%%%%%%%%%%%%%%%%%%%%%%%%%%%%%%%%
%
%         Results
%
%%%%%%%%%%%%%%%%%%%%%%%%%%%%%%%%%%%%%%%%%%%%%%%%%%%%%%%%%%%%%%%%
\section{The 2FHL Catalog}
\label{sec:results}

\begin{figure*}[!ht]
    \centering
    \includegraphics[width=16cm,trim=0 0 0 0cm]{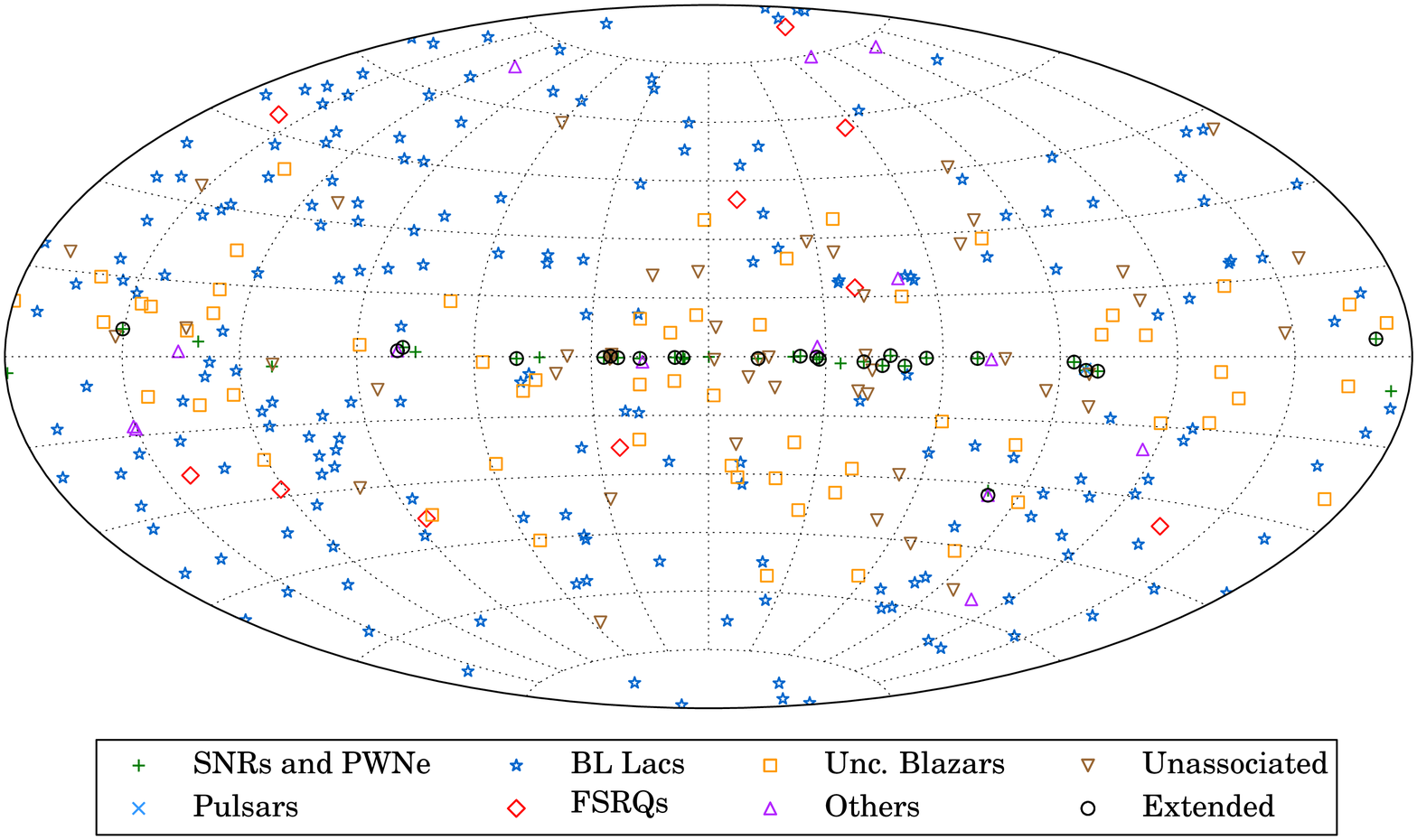} 
    \caption{Sky map, in Galactic coordinates and Hammer-Aitoff projection,
showing the sources in the 2FHL catalog classified by their most likely association. 
    \label{fig:all_sky}}
\end{figure*}

The 2FHL catalog includes 360 sources detected over the whole sky, each with a likelihood test statistic of ${\rm TS}\geq 25$ and number of associated photons, $N_{\rm pred}\geq 3$. 
The association procedure (see \S~\ref{sec:associations}) finds that 75\% of the sources in the catalog (274 sources) are extragalactic\footnote{This includes N~157B, an extragalactic pulsar wind nebula (PWN).}, 11\% (38 sources) are of Galactic nature, and 13\% (48 sources) are unassociated (or associated with a TeV source of unknown nature). The unassociated sources are divided between 23 sources located at $|b|<10^{\circ}$, and 25 sources at $|b|\geq 10^{\circ}$. Therefore the fraction of extragalactic sources in the sample is likely larger than 80\,\%.
{ The number of 2FHL sources that have not been reported in 3FGL is 57, 47 of which have not been previously
reported in any \lat catalog nor in the TeVCat and are thus new $\gamma$-ray sources.}
The results of the association procedures are summarized in Table~\ref{tab:class}. Figure~\ref{fig:all_sky} shows the location of 2FHL sources color-coded according to their source class.

%%%%%%%%%%%%%%%%%%%%%%%%%%%%%%%%%%%%%%%%%%%%%%%%%%%%%%%%%%%%%%%%
%
%         Description of the Catalog
%
%%%%%%%%%%%%%%%%%%%%%%%%%%%%%%%%%%%%%%%%%%%%%%%%%%%%%%%%%%%%%%%%
\subsection{Description of the Catalog}
\label{sec:catalog}

The format of the 2FHL catalog follows that  of previous 
\lat catalogs \citep{1FHL,3FGL} and it is detailed in Table~\ref{tab:description}, while { an excerpt of the catalog itself, which will be fully released in FITS\footnote{See: http://fits.gsfc.nasa.gov/.} format,
 is presented in Table~\ref{tab:catalog}.}
Table~\ref{tab:catalog}  contains 16 columns and it  lists the 2FHL name, position, significance, spectral properties (along with their 1\,$\sigma$ uncertainties), association, { class and redshift (if available)} of the sources. The positions are given in degrees in both Equatorial (J2000) and Galactic coordinates  with their $1\sigma$ positional uncertainties (also in degrees). 
{ The full catalog, which contains 35 columns, also reports (as detailed
in Table~\ref{tab:description}) energy fluxes in the whole 50\,GeV--2\,TeV energy band, and integrated photon fluxes in three logarithmically spaced energy bins together with their 1\,$\sigma$ uncertainties, and the number of photons attributed to the source ($N_{\rm pred}$).  Finally, the most likely association (with probability $>$80\,\%) is also given with its probability. We also report the 3FGL, 1FHL, 1FGL, and TeVCat associations if any.}

\begin{deluxetable}{lcccccccccccclccccc}
\rotate
\setlength{\tabcolsep}{0.04in}
\tablewidth{0pt}
\tabletypesize{\scriptsize}
%\tabletypesize{\tiny}
\tablecaption{Excerpt of the 2FHL Catalog \label{tab:catalog}}
\tablehead{
\colhead{2FHL Name} & 
\colhead{R.A.} &
\colhead{Dec.} &
\colhead{$l$} & 
\colhead{$b$} &
\colhead{$\theta$} &
\colhead{TS} & 
\colhead{$F_{50}$} &
\colhead{$\Delta F_{50}$} &
\colhead{$S_{50}$} & 
\colhead{$\Delta S_{50}$} &
\colhead{$\Gamma$} &
\colhead{$\Delta\Gamma$} &
\colhead{Association} &
\colhead{Class} &
\colhead{Redshift} 
%\colhead{3FGL} &
%\colhead{1FHL} &
%\colhead{TeVCat}
}
\startdata
 J0008.1+4709   &  2.044 &  47.164 &     115.339 & -15.069 & 0.053 &  28.9 &  1.29 &      0.70 &  1.27 &      0.70 &             6.26 &      2.75 & MG4 J000800+4712     & bll        &             2.100   \\%& 3FGL J0008.0+4713 & 1FHL J0007.7+4709 & \nodata        \\
 J0009.3+5031   &  2.343 &  50.522 &     116.124 & -11.793 & 0.035 &  54.5 &  2.00 &      0.82 &  2.12 &      0.92 &             5.07 &      1.65 & NVSS J000922+50302   & bll        &             \nodata \\%& 3FGL J0009.3+5030 & 1FHL J0009.2+5032 & \nodata        \\
 J0018.5+2947   &  4.635 &  29.788 &     114.464 & -32.542 & 0.023 &  31.1 &  1.11 &      0.64 &  2.13 &      1.80 &             2.58 &      0.99 & RBS 0042             & bll        &             0.100   \\%& 3FGL J0018.4+2947 & 1FHL J0018.6+2946 & \nodata        \\
 J0022.0+0006   &  5.500 &   0.106 &     107.172 & -61.862 & 0.042 &  30.3 &  2.05 &      1.00 &  7.15 &      5.50 &             1.86 &      0.57 & 5BZGJ0022+0006       & bll-g      &             0.306   \\%& \nodata           & \nodata           & \nodata        \\
 J0033.6$-$1921 &  8.411 & -19.358 &      94.280 & -81.222 & 0.019 & 149.5 &  5.71 &      1.57 &  7.98 &      2.81 &             3.32 &      0.68 & KUV 00311$-$1938     & bll        &             0.610   \\%& 3FGL J0033.6-1921 & 1FHL J0033.6-1921 & TeV J0033-1921 \\
 J0035.8+5949   &  8.966 &  59.831 &     120.974 &  -2.981 & 0.012 & 380.4 & 12.30 &      1.92 & 30.30 &      7.27 &             2.23 &      0.22 & 1ES 0033+595         & bll        &             \nodata \\%& 3FGL J0035.9+5949 & 1FHL J0035.9+5950 & TeV J0035+5950 \\
 J0040.3+4049   & 10.095 &  40.832 &     120.676 & -21.992 & 0.020 &  27.0 &  1.10 &      0.66 &  2.97 &      2.79 &             2.12 &      0.81 & B3 0037+405          & bcu I      &             \nodata \\%& 3FGL J0040.3+4049 & 1FHL J0040.3+4049 & \nodata        \\
 J0043.9+3424   & 10.976 &  34.411 &     121.164 & -28.435 & 0.051 &  39.9 &  1.91 &      0.86 &  2.13 &      1.04 &             4.56 &      1.60 & GB6 J0043+3426       & fsrq       &             0.966   \\%& 3FGL J0043.8+3425 & 1FHL J0043.7+3425 & \nodata        \\
 J0045.2+2126   & 11.319 &  21.445 &     121.020 & -41.404 & 0.036 &  81.9 &  3.51 &      1.19 &  4.82 &      1.98 &             3.38 &      0.82 & GB6 J0045+2127       & bll        &             \nodata \\%& 3FGL J0045.3+2126 & 1FHL J0045.2+2126 & \nodata        \\
 J0048.0+5449   & 12.002 &  54.828 &     122.433 &  -8.040 & 0.049 &  35.4 &  1.27 &      0.64 &  7.99 &      5.97 &             1.30 &      0.51 & 1RXS J004754.5+544   & bcu II     &             \nodata \\%& 3FGL J0047.9+5447 & \nodata           & \nodata        \\
\enddata
\tablecomments{R.A. and Dec. are Equatorial coordinates in J2000 epoch, $l$ and $b$ are Galactic coordinates. All coordinates are shown in degrees. $\theta$ is the 68\,\% uncertainty radius. TS is the test statistic. $F_{50}$ and $\Delta F_{50}$ are the integrated photon flux between 50~GeV and 2~TeV and its uncertainty in units of $10^{-11}$~photon~cm$^{-2}$~s$^{-1}$. $S_{50}$ and $\Delta S_{50}$ are the energy flux between 50~GeV and 2~TeV and its uncertainty in units of $10^{-12}$~erg~cm$^{-2}$~s$^{-1}$. $\Gamma$ and $\Delta \Gamma$ are the photon  index and its uncertainty from a power-law fit. All the uncertainties are $1\sigma$ uncertainties unless stated otherwise. See text for details on the association methodology. The source classes are detailed in Table~\ref{tab:class}. Redshifts were taken from from \cite{shaw12}, \cite{shaw13}, \cite{masetti13} and the NED and SIMBAD databases.}
\end{deluxetable}

%%%%%%%%%%%%%%%%%%%%%%%%%%%%%%%%%%%%%%%%%%%%%%%%%%%%%%%%%%%%%%%%
%
%         General Results
%
%%%%%%%%%%%%%%%%%%%%%%%%%%%%%%%%%%%%%%%%%%%%%%%%%%%%%%%%%%%%%%%%
\subsection{General Characteristics of 2FHL Sources}

\begin{figure}[!ht]
\begin{centering}
     \includegraphics[width=\columnwidth]{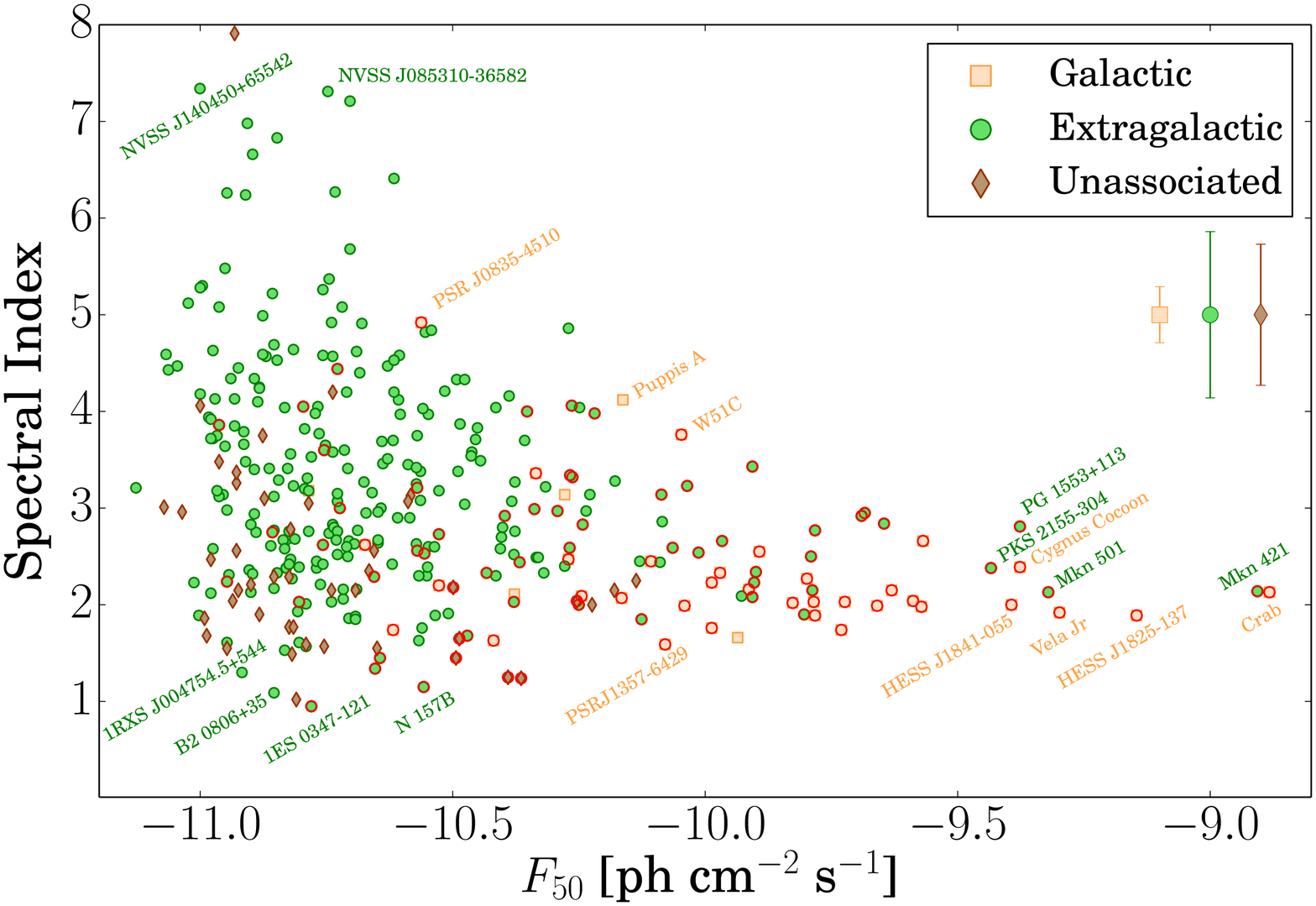}
    \caption{The photon flux versus the spectral index of Galactic sources (orange squares), extragalactic sources (green circles), and unassociated sources (brown diamonds). The { medians} of the uncertainties are shown as well. We note that the detectability does not significantly depend on the spectral index as a consequence of the low intensity of the diffuse background and constant PSF over the energy range of the analysis. { Symbols with a red outline are sources already detected at TeV energies and contained in the TeVCat catalog.}
    \label{fig:index_vs_flux}}
\end{centering}
\end{figure}

The 2FHL sources have $>$50\,GeV fluxes ranging from
$\sim$$8\times 10^{-12}$~ph~cm$^{-2}$~s$^{-1}$ to $\sim$$1.3\times 10^{-9}$~ph~cm$^{-2}$~s$^{-1}$
with a median flux of  $2.0\times 10^{-11}$~ph~cm$^{-2}$~s$^{-1}$
and a median spectral index of 2.83. The index uncertainty increases rapidly with the spectral index (\eg the uncertainty is about
$\pm 0.5$ for sources with $\Gamma=2$ whereas it is 
 $\pm 2$ for sources with $\Gamma=5$).
Half of the sources are localized to better than
1.7$'$ radius at 68\,\% confidence. 
Figure~\ref{fig:index_vs_flux} plots the spectral index versus the photon flux for sources associated with extragalactic sources or located at \blot (the extragalactic sample), Galactic sources, and unassociated sources. Figure~\ref{fig:index_vs_flux} shows that there is no visible dependence of the sensitivity (i.e. minimum detectable photon flux) on the spectral index. This is because the size of the Pass~8 PSF remains approximately constant above 50\,GeV. { However, extragalactic sources are detected to lower fluxes than Galactic objects,
 highlighting that the sensitivity for source detection becomes { worse} in the plane of the Galaxy.}

The distributions of spectral indices and the highest photon energy reported in Figure~\ref{fig:hist_index} show that extragalactic sources tend to have larger photon indices (median of 3.13) than Galactic sources (median of 2.10).   Because of the harder spectra, Galactic sources tend to have higher-energy HEPs than those of extragalactic sources
as shown as well in Figure~\ref{fig:hist_index}.
It is interesting to note that unassociated sources have a median index of 2.22 (2.00 for  sources at \blat and 2.96 for those at \blot), showing that a fraction (see later) of unassociated sources is likely of Galactic origin. 

Building a spectral energy distribution (SED) represents a powerful way to discriminate or infer the nature of a source. By combining the spectral data from the 3FGL, 1FHL, and 2FHL catalogs, it becomes possible to measure the SEDs
of sources over four decades in energy.
Although these catalogs rely on different exposures and  most $\gamma$-ray sources are variable, these data allow us to characterize the high-energy peak of their broadband SEDs. The SEDs of a few notable sources will be shown in the next sections.

%%%%%%%%%%%%%%%%%%%%%%%%%%%%%%%%%%%%%%%%%%%%%%%%%%%%%%%%%%%%%%%%
%
%         Galactic Science
%
%%%%%%%%%%%%%%%%%%%%%%%%%%%%%%%%%%%%%%%%%%%%%%%%%%%%%%%%%%%%%%%%

\subsection{The 2FHL Galactic Source Population}

The narrow PSF core (about $0.1^{\circ}$) and moderate Galactic diffuse emission (in comparison with the $>100$\,MeV band) allows the LAT to characterize and study well the emission of sources in the plane of our Galaxy above 50\,GeV. Within $|b|<10^{\circ}$, \lat has detected 103 sources.
Of those, 38 sources are associated with Galactic sources, 42 to blazars,
14 are unassociated and 9 are associated with other $\gamma$-ray sources
whose origin is not known (see below).
Figure~\ref{fig:gp1} shows cut-outs of the Galactic plane with all detected sources labeled.

\begin{figure*}[!ht]
\begin{centering}
    \includegraphics[width=8cm]{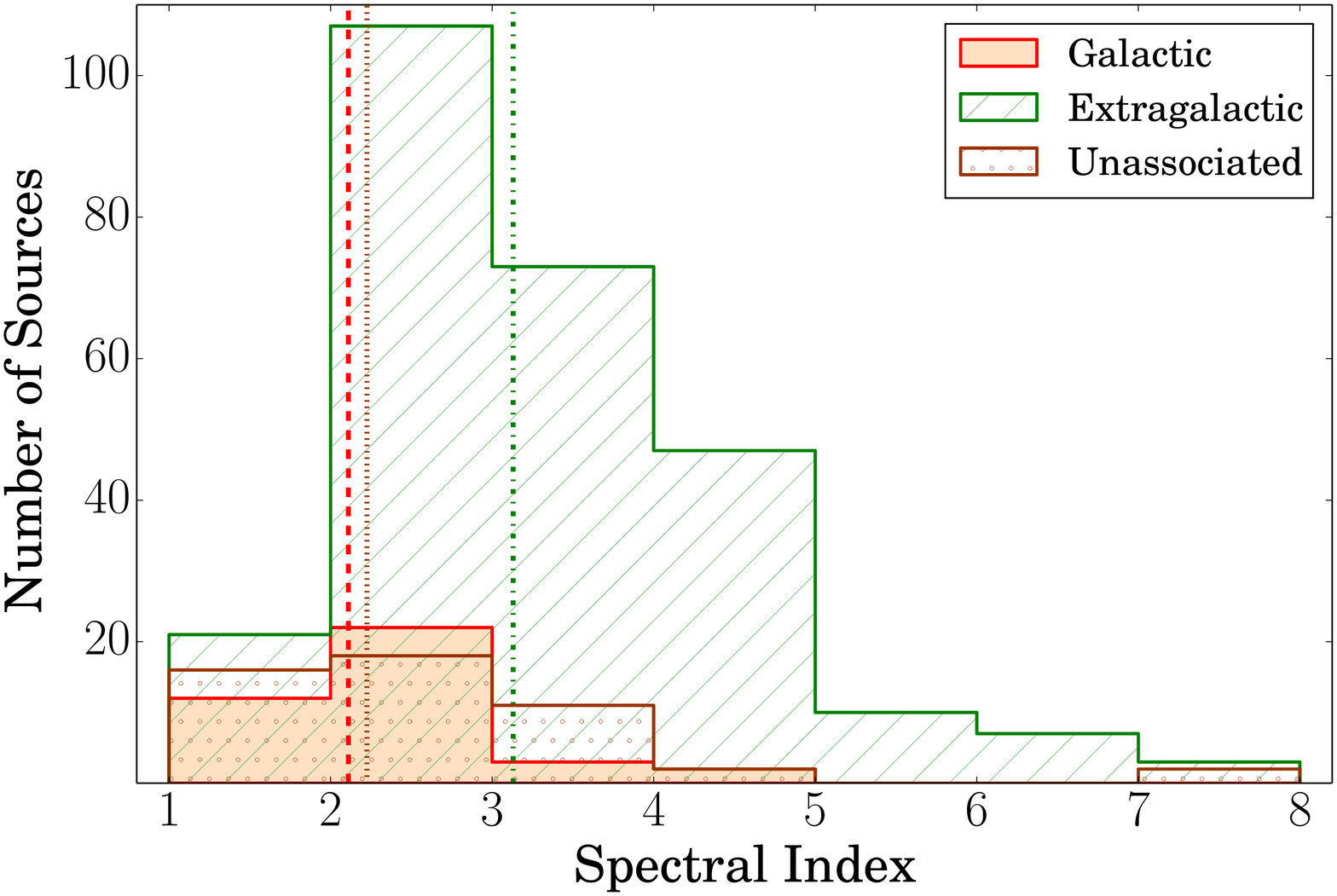}
    \includegraphics[width=8cm]{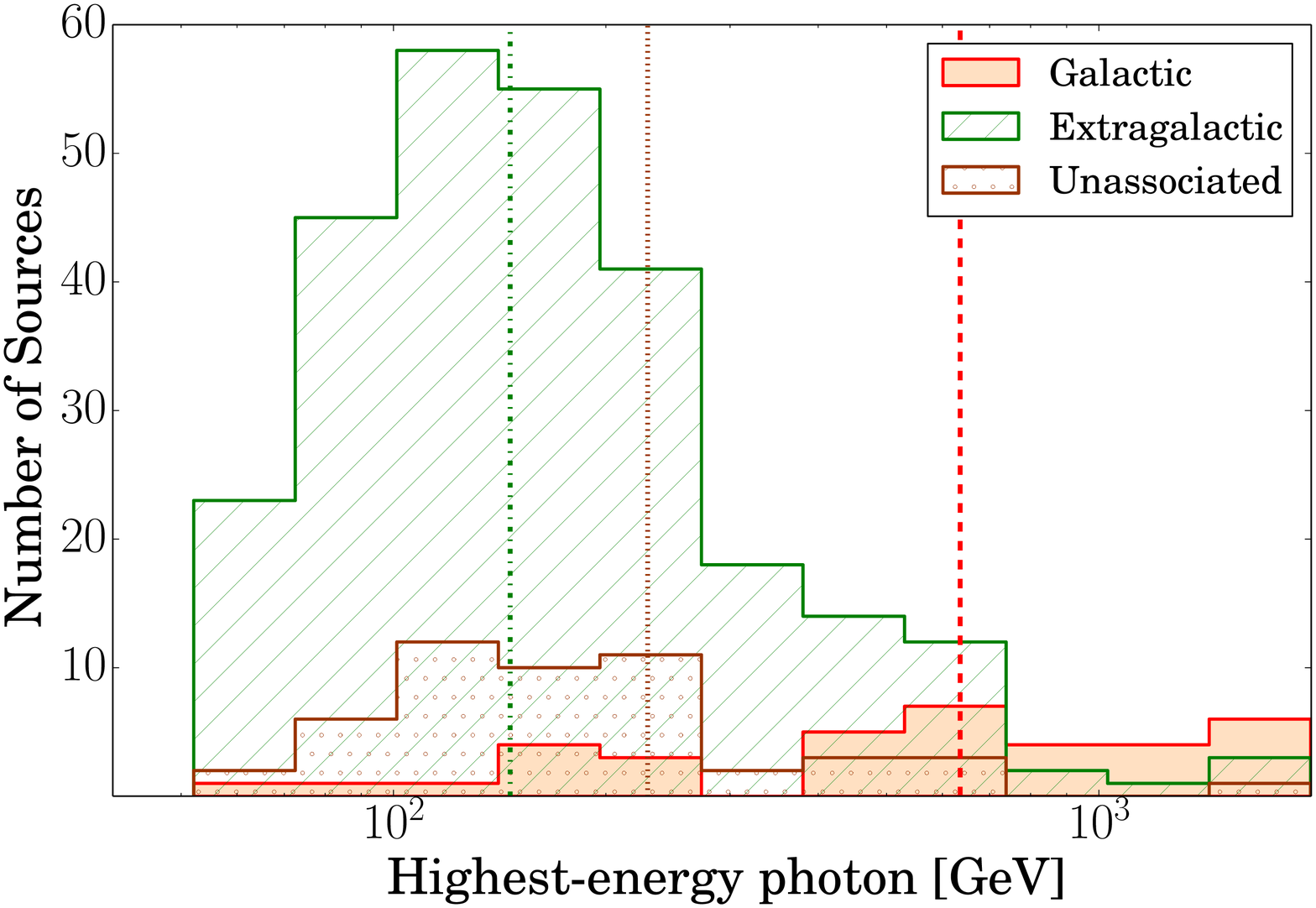} 
    \caption{Distribution of the spectral indices ({\it left panel}) and highest photon energy ({\it right panel}) of the Galactic sources (orange), extragalactic sources (green slash), and unassociated sources (brown dotted). The medians of the distributions are plotted with dashed, dash-dotted, and dotted vertical lines, respectively. Both plots show that a distinct population of hard-spectrum sources is of Galactic origin.
    \label{fig:hist_index}}
\end{centering}
\end{figure*}

Among the 38 Galactic sources, 16 are spatially coincident with SNRs, 13 are coincident with PWNe, 4 are associated with PWN/SNR complexes
and the other 5 sources are X-ray binaries (3), one pulsar (PSR~J0835$-$4510)
and the Cygnus Cocoon.
It is clear that the majority of Galactic sources detected above 50\,GeV
are associated with objects at the final stage of stellar evolution.

\begin{figure*}[!ht]
\begin{center}
\begin{tabular}{ll}
        \includegraphics[angle=90,scale=.3]{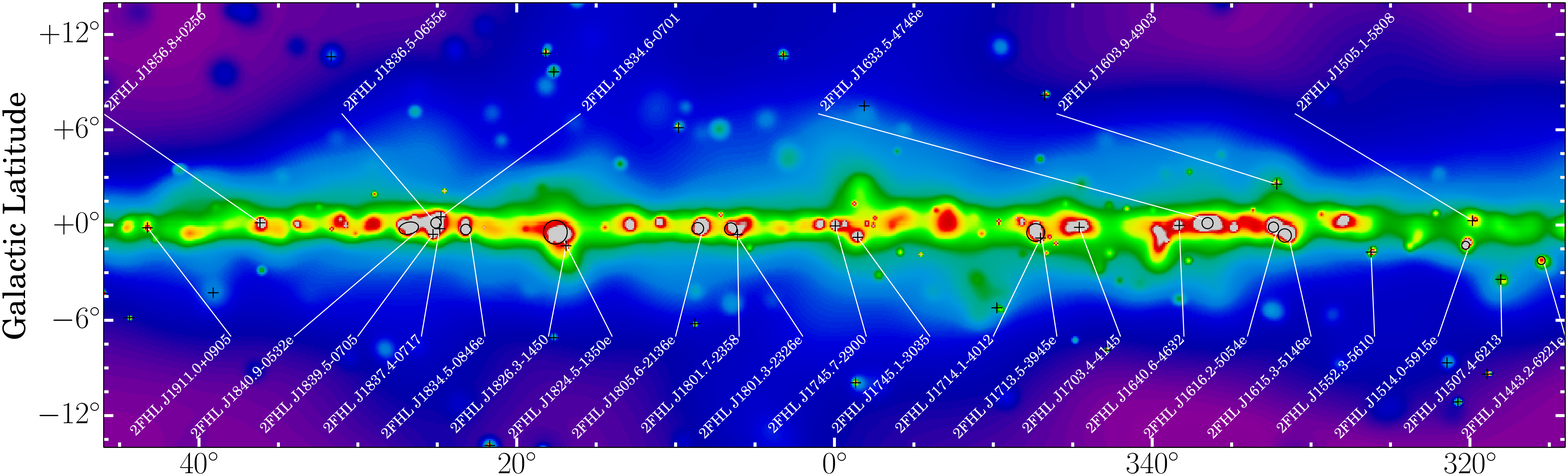}&
        \includegraphics[angle=90,scale=.3]{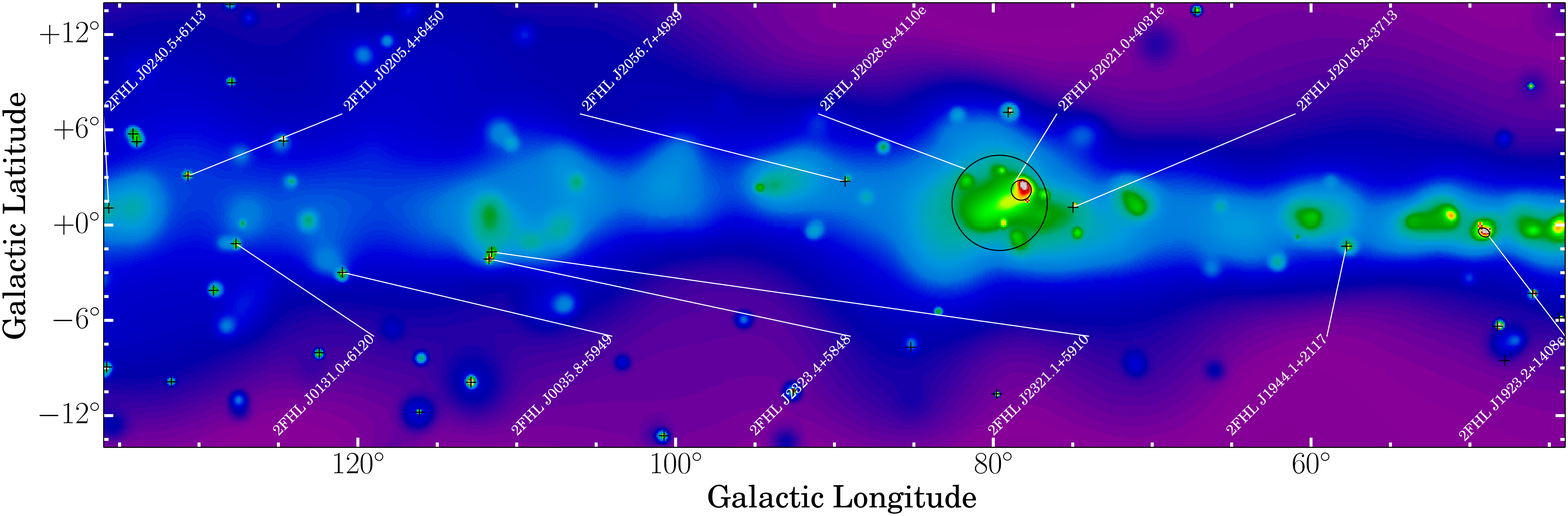}\\
\end{tabular}
\end{center}
    \caption{Adaptively smoothed count map showing the whole Galactic plane $0^{\circ}\leq l\leq 360^{\circ}$ at Galactic latitudes $-14^{\circ} \leq b\leq 14^{\circ}$ divided in four  panels. The panels are centered at $l=0^{\circ}$, $90^{\circ}$, $180^{\circ}$ and $270^{\circ}$, respectively. Detected point sources are marked with a cross whereas extended sources are indicated with  their extensions. Only sources located at $-4^{\circ} \leq b\leq 4^{\circ}$ are explicitly named, plus the Crab Nebula.
    \label{fig:gp1}}
\end{figure*}

\begin{figure*}[!ht]
\begin{center}
\begin{tabular}{ll}
       \includegraphics[angle=90,scale=.3]{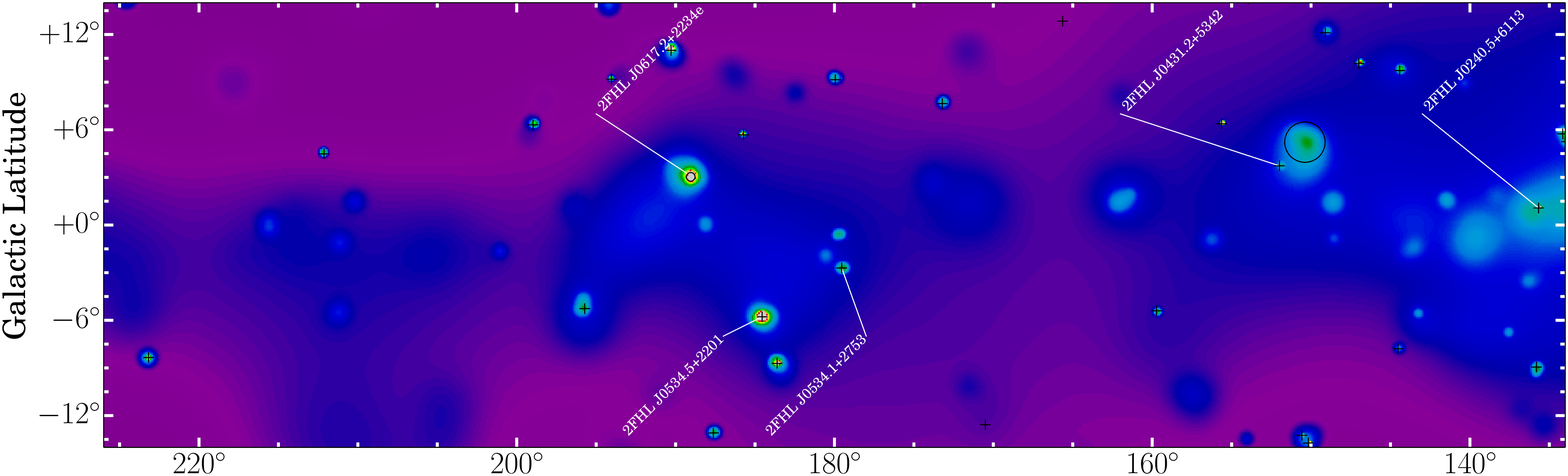}&
        \includegraphics[angle=90,scale=.3]{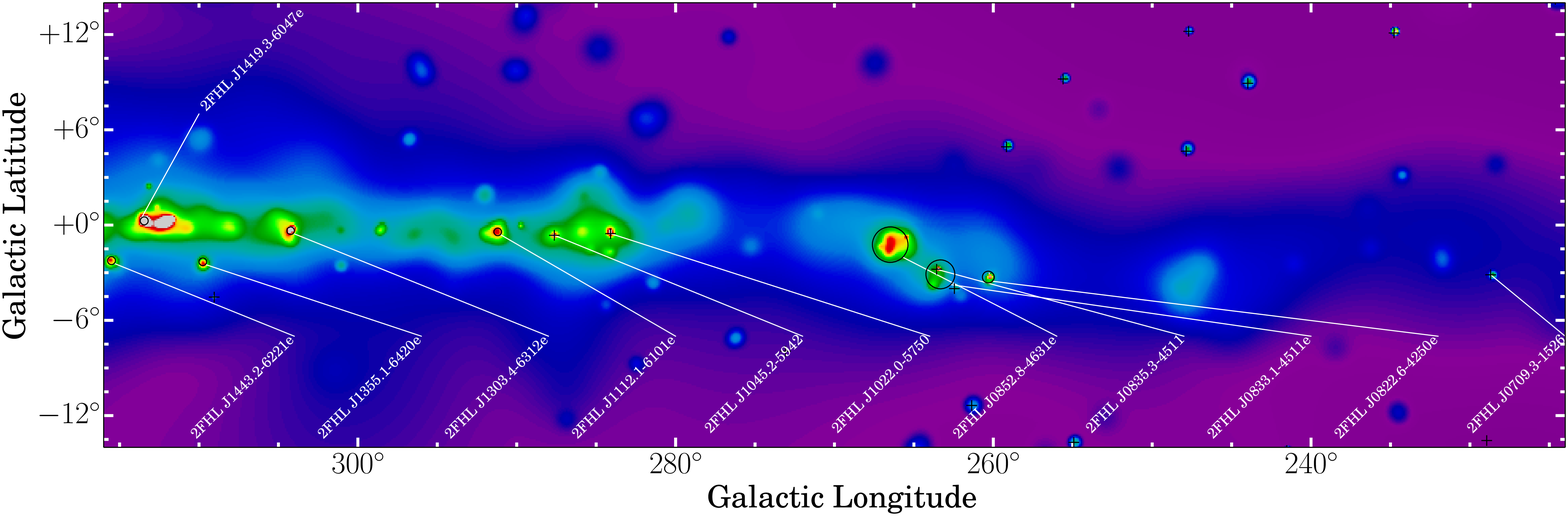}\\
\end{tabular}
\end{center}
    \begin{flushleft}
    {Fig.~\ref{fig:gp1}.}---  continued
    \end{flushleft}
\end{figure*}

Galactic sources display on average hard spectra, which is a sign
of efficient particle acceleration. Roughly 55\% of all Galactic
sources have a spectral index lower than 2.2. For comparison, only 14\% of the { 2FHL} blazars
display such hard spectra. 
 A sizable fraction (approximately 25\%, 
see Figure~\ref{fig:hist_index}, left panel)
of Galactic sources has a photon
index harder than 2, implying a high-energy SED peak in the TeV band.
Indeed, as the right panel of Figure~\ref{fig:hist_index} shows,
\lat detects emission from many Galactic sources well beyond 500\,GeV.
All PWNe detected by {\it Fermi} are found to be powered by young
and energetic pulsars \citep[age $\lesssim 30$\,kyr,][]{Acero13}.
While it is common for PWNe to show hard spectra, this is less
so for SNRs whose majority (about 85\,\%) display softer spectra
\citep{hewittSNRcat13}. Hard-spectrum SNRs are typically young or 
mid-aged ($\lesssim$3--5\,kyr)
and might be difficult to find in radio surveys. Thus, Galactic surveys
at above 50\,GeV have the capability to detect new SNRs that
might have been previously missed.
Such an example is represented by the extended source 
2FHL J0431.2+5553e which is spatially coincident with a 
new SNR (SNR G150.3+4.5) recently reported by \cite{Gao14}.

Of the 14 sources at $|b|<10^{\circ}$ that do not have an
association, 7 have power-law indices harder than 2
which renders them  likely Galactic objects.
It is interesting to note that { 6 of these 7 objects} are offset from the plane of the Galaxy  by more than $4^{\circ}$. This is in marked contrast with the 
associated portion of the sample where only the Crab Nebula and the
newly discovered SNR G150.3+4.5 (out of 34 SNR/PWN systems) have such
a large offset. Thus it seems unlikely that all these unassociated sources
are SNR/PWN systems.

%%%%%%%%%%%%%%%%%%%%%%%%%%%%%%%%%%%%%%%%%%%%%%%%%%%%%%%%%%
%
% H E S S 
%
%%%%%%%%%%%%%%%%%%%%%%%%%%%%%%%%%%%%%%%%%%%%%%%%%%%%%%%%%%%

\subsubsection{Comparison with the H.E.S.S. Galactic Plane Survey}
The H.E.S.S array, with a field of view of about 5$^{\circ}$ and an angular resolution of approximately 0.12$^\circ$, has invested 2800\,hrs of exposure to survey  part\footnote{The H.E.S.S. Galactic plane survey extends between 283$^{\circ}<l<$59$^{\circ}$ and Galactic latitudes of $|b|<3.5^{\circ}$.} 
of the Galactic plane, reaching an average sensitivity of 2\,\% of the Crab Nebula flux (i.e. 4.5$\times10^{-13}$\,ph~cm$^{-2}$~s$^{-1}$) at $\geq$1\,TeV \citep{aharonian06_gps,carrigan2013}. Considering that the Crab Nebula spectrum is harder
in the 2FHL band than in the $>$1\,TeV band, we estimate that the average sensitivity of 2FHL in the same region of the H.E.S.S. survey is $\sim$3--4\,\% of the { 50\,GeV--2\,TeV Crab Nebula flux.} The slightly better sensitivity 
allows H.E.S.S. to detect 69 sources (as reported in the TeVCat), while
the LAT finds 36 objects in the same area. However, the comparable sensitivities of the two surveys allow the study of the  properties of the high-energy Galactic population.
In the 2FHL catalog there is almost an equal number of SNRs and PWNe
in contrast to what is found in the  H.E.S.S. survey where the ratio
of PWNe to SNRs is 1.5 to 1. This might be because
the hardest PWNe and softest SNRs { are difficult to detect} respectively
in the $>$50\,GeV and $>$1\,TeV bands.

\begin{figure*}[ht]
\begin{center}
\begin{tabular}{ll}
\includegraphics[width=8cm]{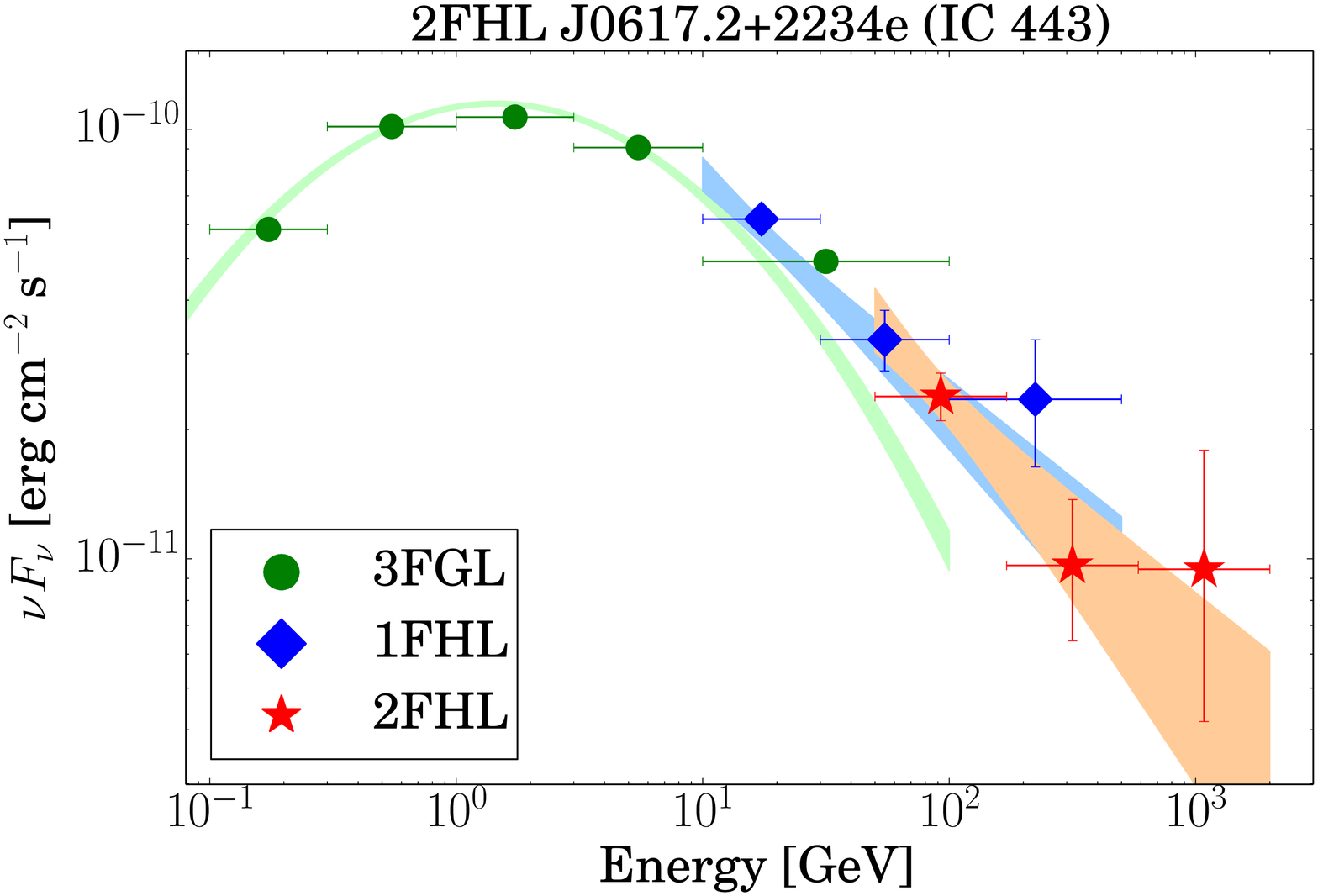} &
\includegraphics[width=8cm]{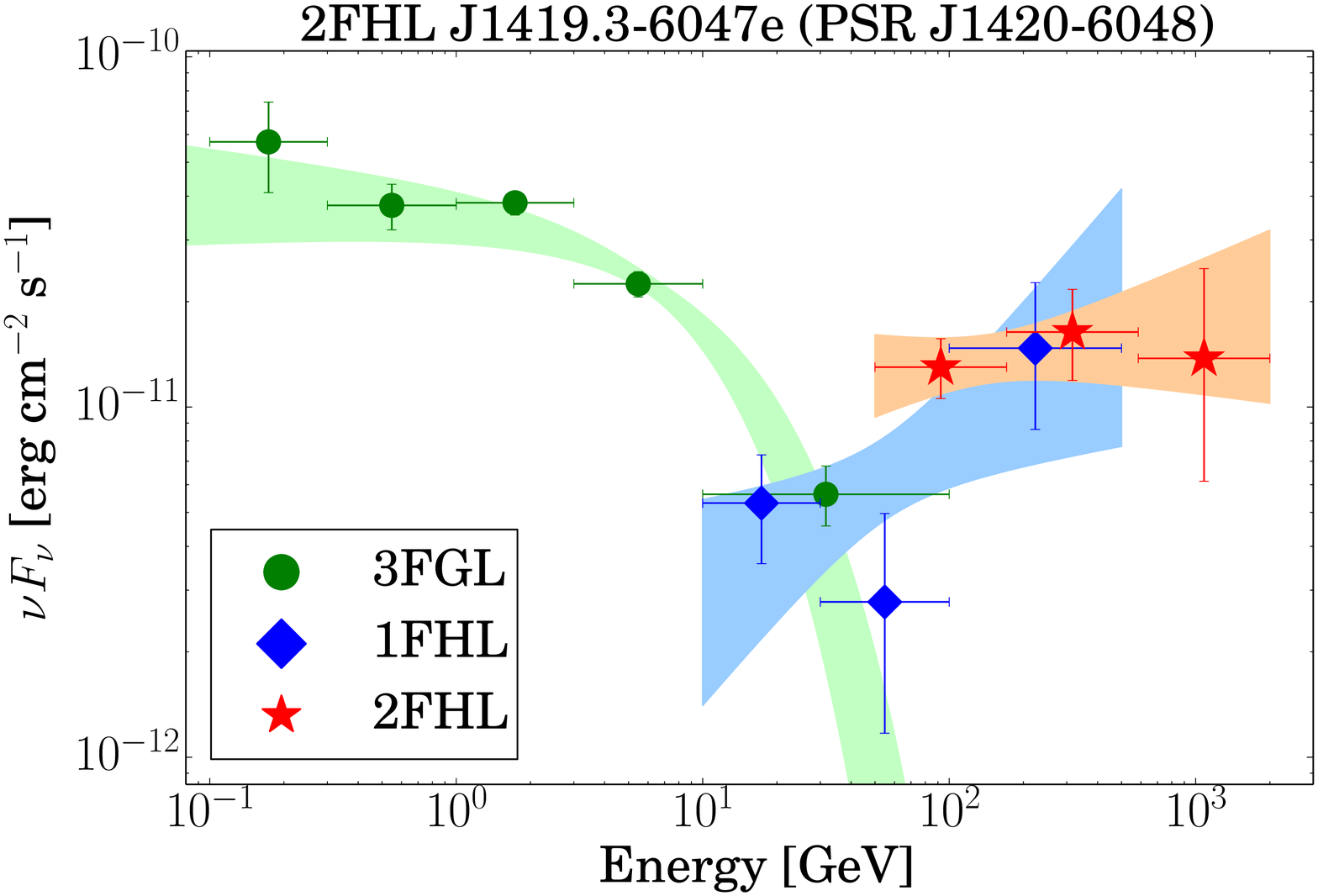}\\
\includegraphics[width=8cm]{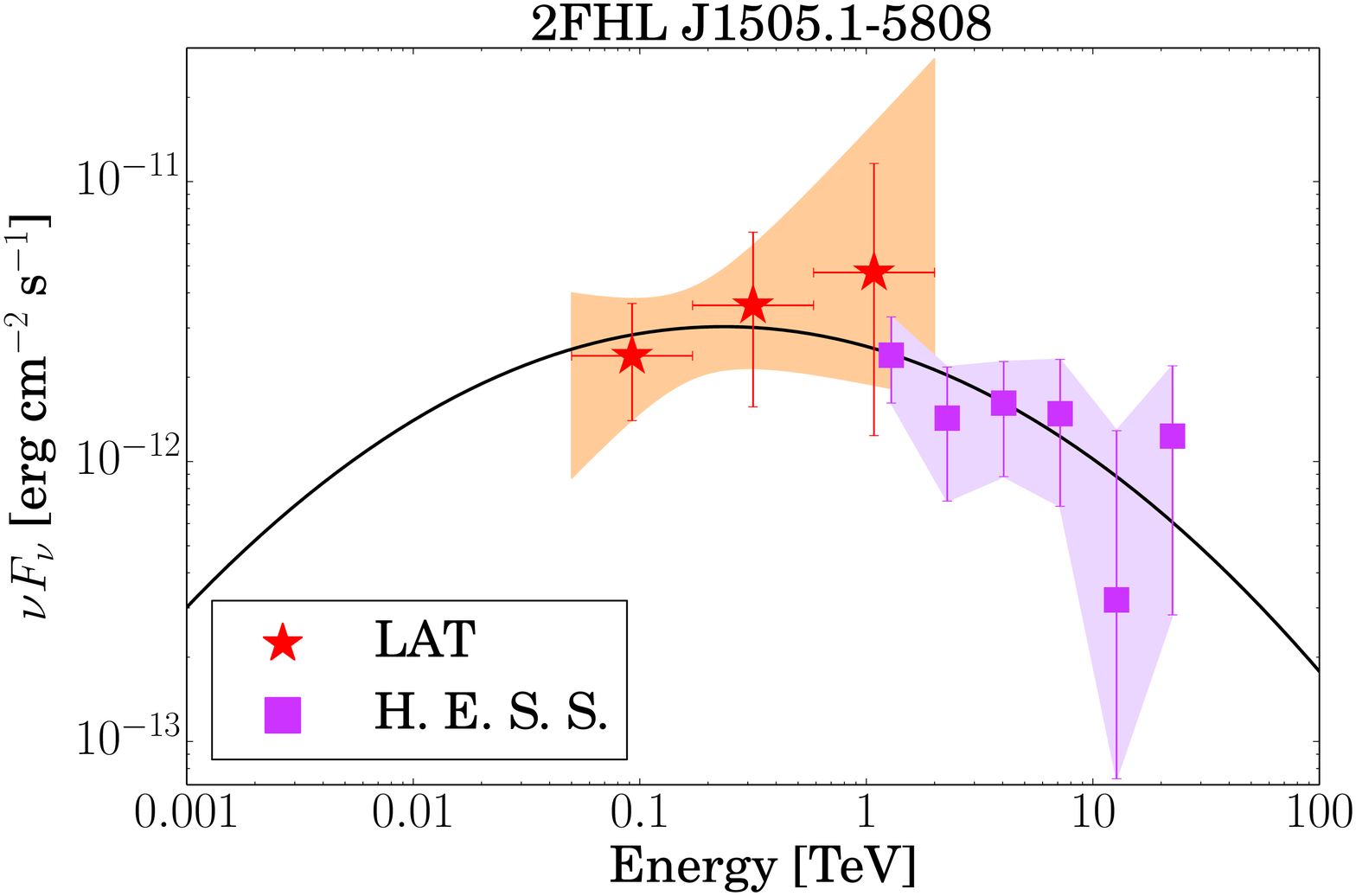} &
\includegraphics[width=8cm]{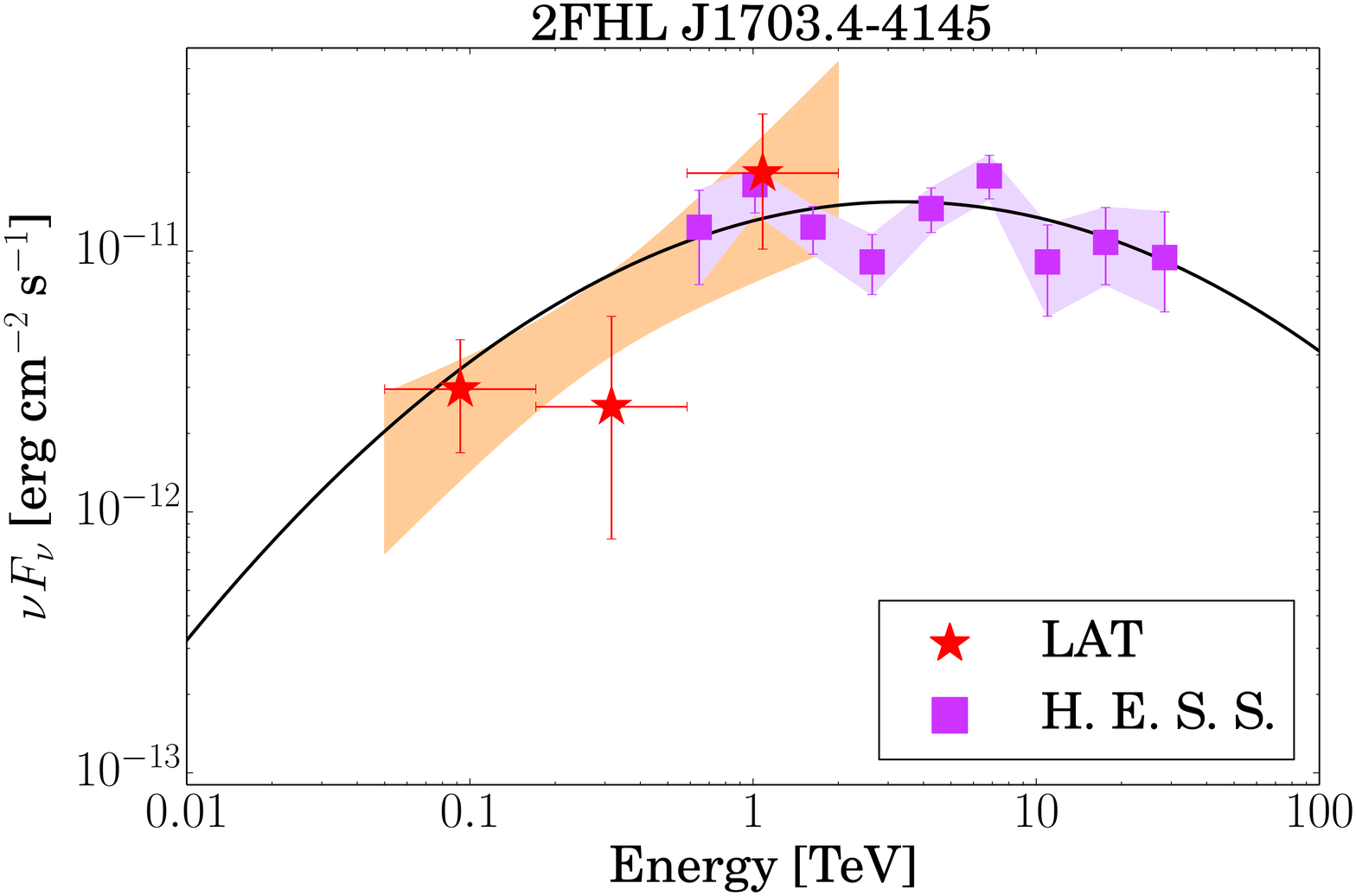} \\
\end{tabular}
\end{center}
\caption{
\label{fig:gal_sed}Spectral energy distributions of four Galactic sources constructed by combining data from the 3FGL (green diamonds), 1FHL (blue circles), and 2FHL (red stars). We show the 3FGL extended source SNR IC~443 (\emph{top left}), the new 2FHL extended source PSR~J1420$-$6048 (\emph{top right}), and two ``dark accelerators'' detected by H.E.S.S. at TeV energies \citep[][purple squares]{carrigan2013} without a previous LAT counterpart:  { HESS~J1503$-$582} (\emph{bottom left}) and HESS~J1702$-$420 (\emph{bottom right}).}
\end{figure*}

Of the 36 2FHL sources that fall within the footprint of 
the H.E.S.S. survey, 23 have already been detected
at TeV energies and are associated with known counterparts,
while 7 are undetected. The remaining 6 objects
(2FHL~J1022.0$-$5750, 2FHL~J1505.1$-$5808,  2FHL~J1507.4$-$6213, 2FHL~J1703.4$-$4145, 2FHL~J1745.1$-$3035 and  2FHL~J1856.8+0256)
are spatially coincident with TeV sources whose origin is not known.
All of them have hard spectral indices ($\Gamma<$2.2), but
it is interesting to note that 4 of them 
(2FHL~J1022.0$-$5750, 2FHL~J1505.1$-$5808, 2FHL~J1703.4$-$4145, and 2FHL~J1745.1$-$3035)
have $\Gamma<1.7$ (see also Figure~\ref{fig:gal_sed}).

We find that 2FHL~J1022.0$-$5750 is spatially compatible with
{ HESS~J1023$-$575, an extended TeV source \citep{westerlund2_hess11},
whose emission might be due to a PWN powered by PSR~J1023$-$5746 \citep{Acero13}. }
2FHL~J1505.1$-$5808 is spatially coincident with the unidentified
object HESS~J1503$-$582, which has a size of 0.26$^{\circ}$
{ and a flux above 1\,TeV \citep{renaud08} 
compatible with the extrapolation of the 2FHL J1505.1$-$5808 spectrum.}
Its spectrum, reminiscent of that of a PWN 
\citep[\eg, HESS~J1825$-$137,][]{grondin2011} is reported in Figure~\ref{fig:gal_sed}.

2FHL~J1507.4$-$6213 is spatially coincident with HESS~J1507$-$622, an extended source with a radius of 0.15$^{\circ}$ located  3.5$^{\circ}$ from the plane \citep{acero11}. The analysis of multiwavelength data showed that it is not possible to discriminate between a hadronic and leptonic origin of the emission, but that the latter scenario, if the emission is powered by a PWN, would require a pulsar generated in the explosion of a hyper-velocity star in order to reach the required distance from the plane \citep{domainko2012}.

The sources 2FHL~J1703.4$-$4145 and 2FHL~J1745.1$-$3035 are the hardest sources ($\Gamma<1.3$) among the six objects.
2FHL~J1703.4$-$4145 is spatially coincident with the bright radio emission
observed from the western side of the shell of SNR G344.7$-$001,
a nearby mid-aged  shell-type (age $\sim3000$\,yr and 8$'$ diameter) SNR \citep{giacani2011}. Both the 2FHL source and the SNR are spatially coincident
with the larger, elongated and unidentified HESS~J1702$-$420 \citep{aharonian08}.
It thus seems likely that  SNR G344.7$-$001 is the  counterpart
of  2FHL J1703.4$-$4145 and perhaps also of HESS~J1702$-$420.
The combined {\it Fermi}-H.E.S.S. spectrum of this source is reported in Figure~\ref{fig:gal_sed}.

2FHL~J1745.1$-$3035 is found to be spatially coincident with 
the extended source HESS~J1745$-$303, which may be comprised of
up to three different sources \citep{aharonian2008_j1745}. Indeed, the position of
2FHL~J1745.1$-$3035 is compatible  with the 'C' emission
region \citep[the second brightest region in the complex,][]{aharonian2008_j1745}.
However, the nature of this source is more complex, because
the 2FHL source is marginally brighter at 1\,TeV than the entire H.E.S.S. region and has also a harder spectrum (spectral index of 1.25$\pm0.38$ in 2FHL 
versus $2.17\pm 0.11$ as measured by H.E.S.S.).

Finally, 2FHL~J1856.8+0256 is coincident with HESS~J1857+026, an almost radially symmetric extended source  \citep{aharonian08_unid}, whose emission likely originates from a PWN powered by PSR~J1856+0245 \citep{rosseau2012}.

%%%%%%%%%%%%%%%%%%%%%%%%%%%%%%%%%%%%%%%%%%%%%%%%%%%%%%%%%%%%%%%%%%%
%
%  Extended Source Results
%
%%%%%%%%%%%%%%%%%%%%%%%%%%%%%%%%%%%%%%%%%%%%%%%%%%%%%%%%%%%%%%%

\subsubsection{\label{sec:ESresults}Extended Source Results}

In total, 31 sources are modeled as spatially extended and input into the ML analysis: 25 listed in 3FGL, 5 sources detected in the {\tt pointlike} analysis (described in $\S$ \ref{sec:3FGL_ES}) that were not { detected as extended at the time of} 3FGL, and one, SNR W41, reported  recently by both the H.E.S.S. and LAT teams \citep{HESSLATW41}. Names and properties of the extended sources  are provided in Tables \ref{tab:extended} and \ref{tab:new_extended}. 
Six extended sources, detected in 3FGL, were not detected in 2FHL: the SMC, S~147 ({the point source 2FHL~J0534.1+2753 was detected inside it}), the lobes of Centaurus A (although we detect its core as a point source, 2FHL J1325.6$-$4301), W~44, HB~21 and the Cygnus Loop.

We detect a weak source, 2FHL~J1714.1$-$4012 (TS = 27), just outside the southwestern edge of the 3FGL spatial template used to model the emission from SNR RX J1713.7$-$3946 (2FHL~J1713.5$-$3945e). 2FHL~J1714.1$-$4012 has a hard spectral index $\Gamma = 1.63 \pm 0.38$, that is within errors of the spectral index derived for the SNR, $\Gamma = 2.03 \pm 0.20$. It is unclear whether 2FHL~J1714.1$-$4012 is a distinct source separated from the SNR, or the result of un-modeled residual emission due to an imperfection in the spatial template adopted for the extended source.

2FHL~J1836.5$-$0655e is associated with the PWN HESS J1837$-$069. The 3FGL catalog contains  several point sources in the vicinity of the PWN. We detect three sources in the vicinity, 2FHL~J1834.5$-$0701, 2FHL~J1837.4$-$0717 and 2FHL~J1839.5$-$0705, the first two of which are coincident with 3FGL sources (3FGL J1834.6$-$0659, 3FGL J1837.6$-$0717 respectively). The power-law spectral indices of the three 2FHL point sources and 2FHL J1836.5$-$0655e are all consistent with each other. The concentration of sources around HESS J1837$-$069 combined with the spectral compatibility of the sources is suggestive of a common origin to the $\gamma$-ray emission in this region. However, the surrounding $\gamma$ rays could arise from other sources in the region \citep{Gotthelf08}; further analysis is necessary to determine the nature of the sources in this region. 

\begin{figure*}[!ht]
\begin{centering}
    \vspace*{-1cm}
    \includegraphics[width=8cm]{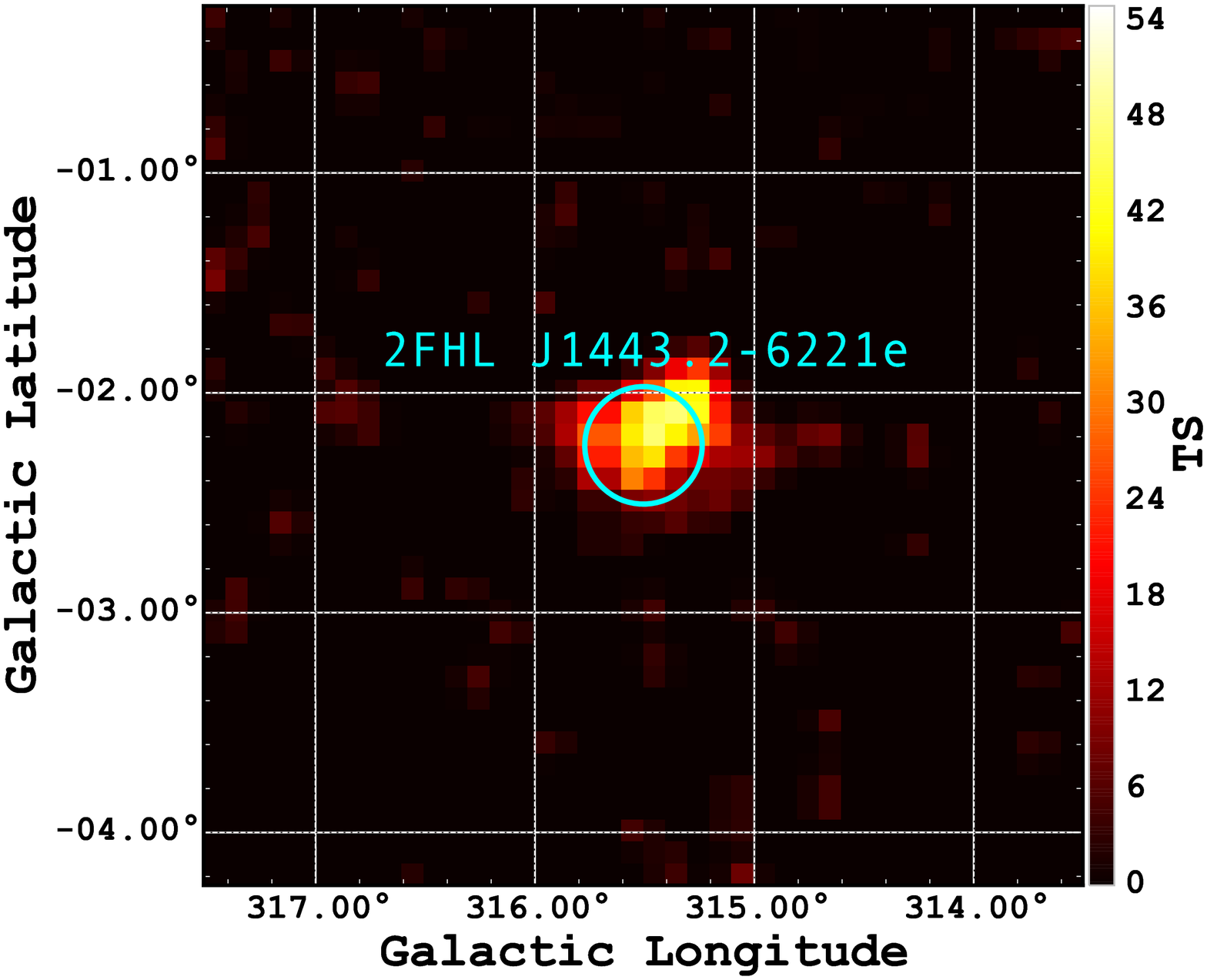}
    \includegraphics[width=8cm]{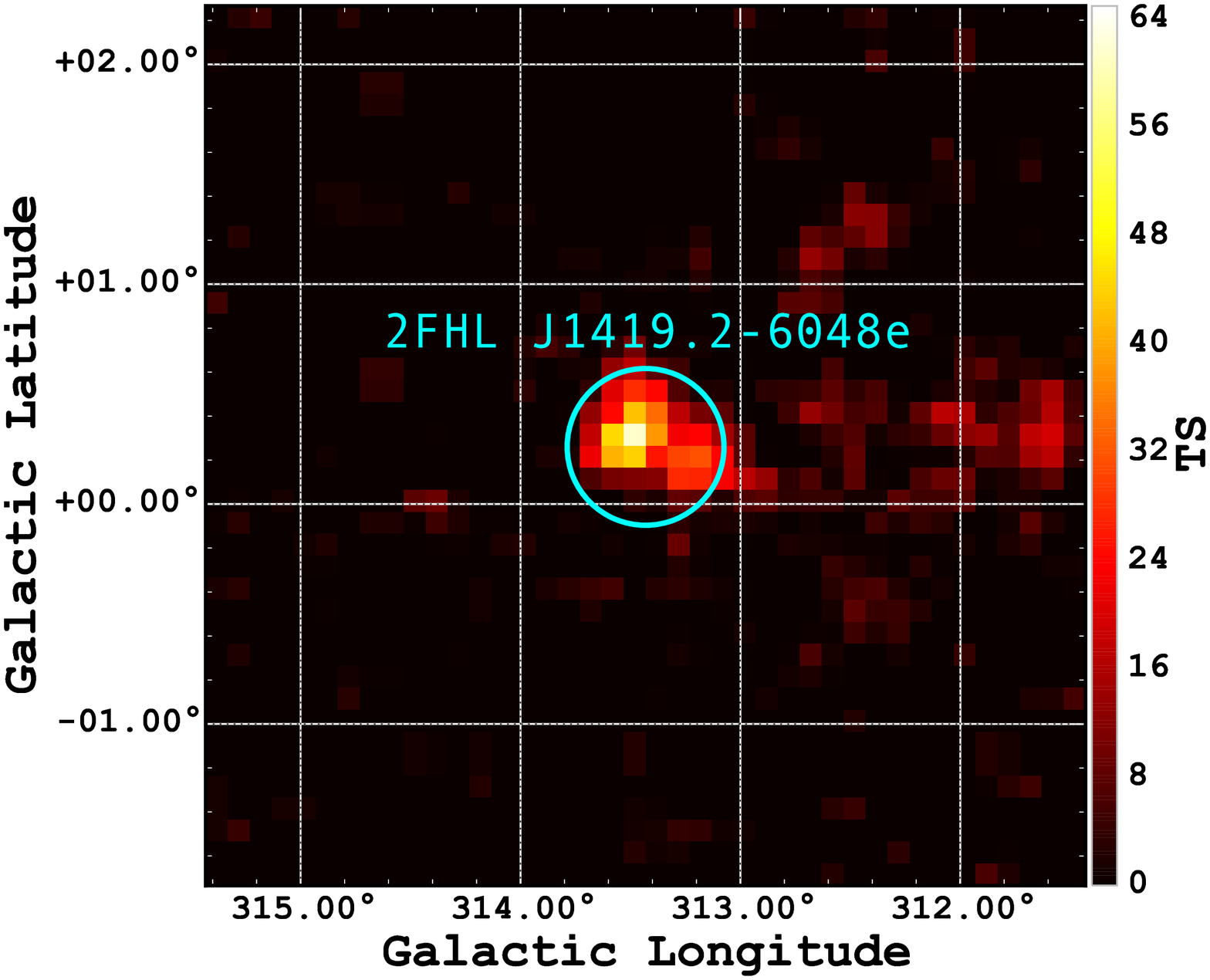}
    \includegraphics[width=8cm]{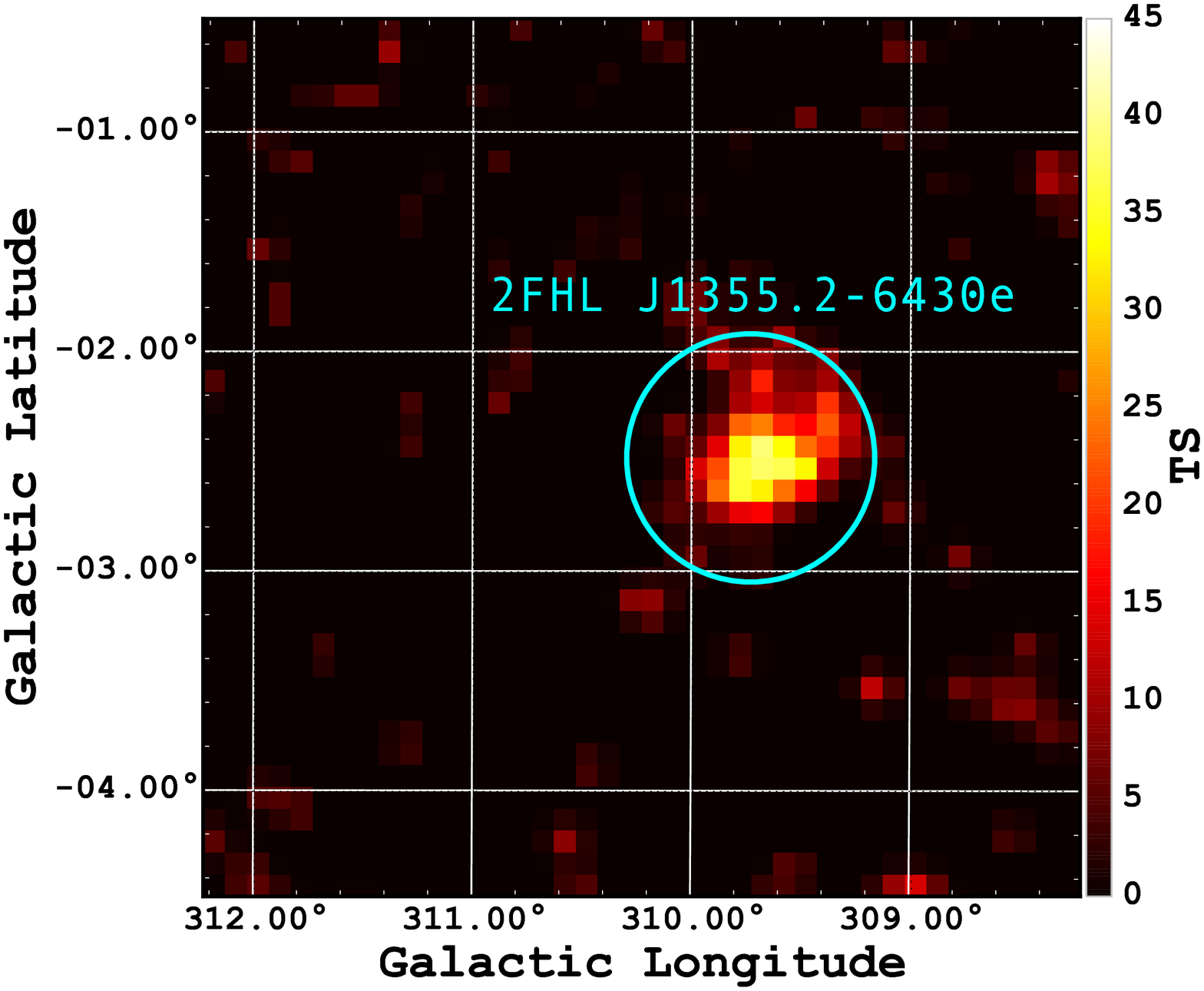}
    \includegraphics[width=8cm]{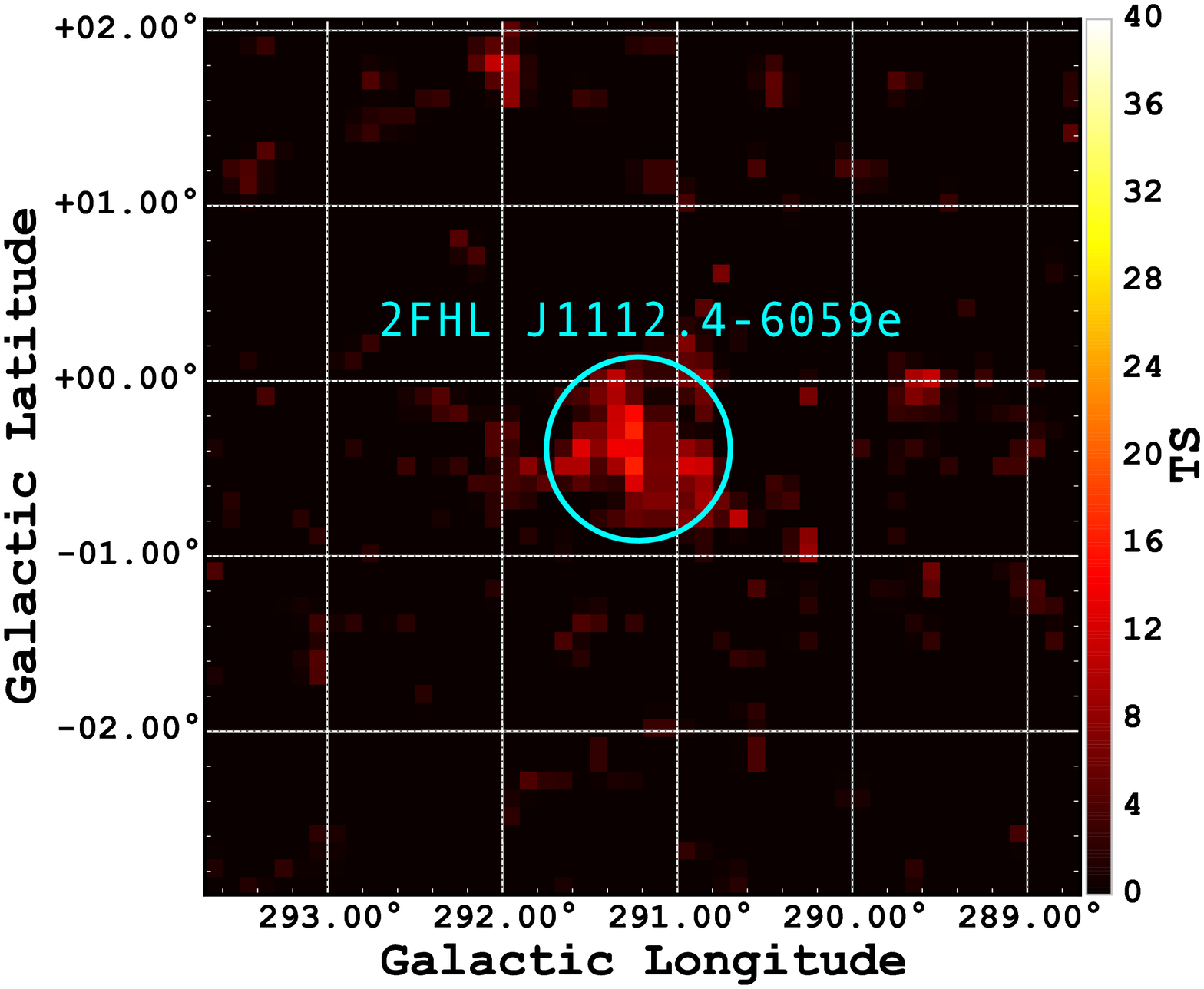}
    \includegraphics[width=8cm]{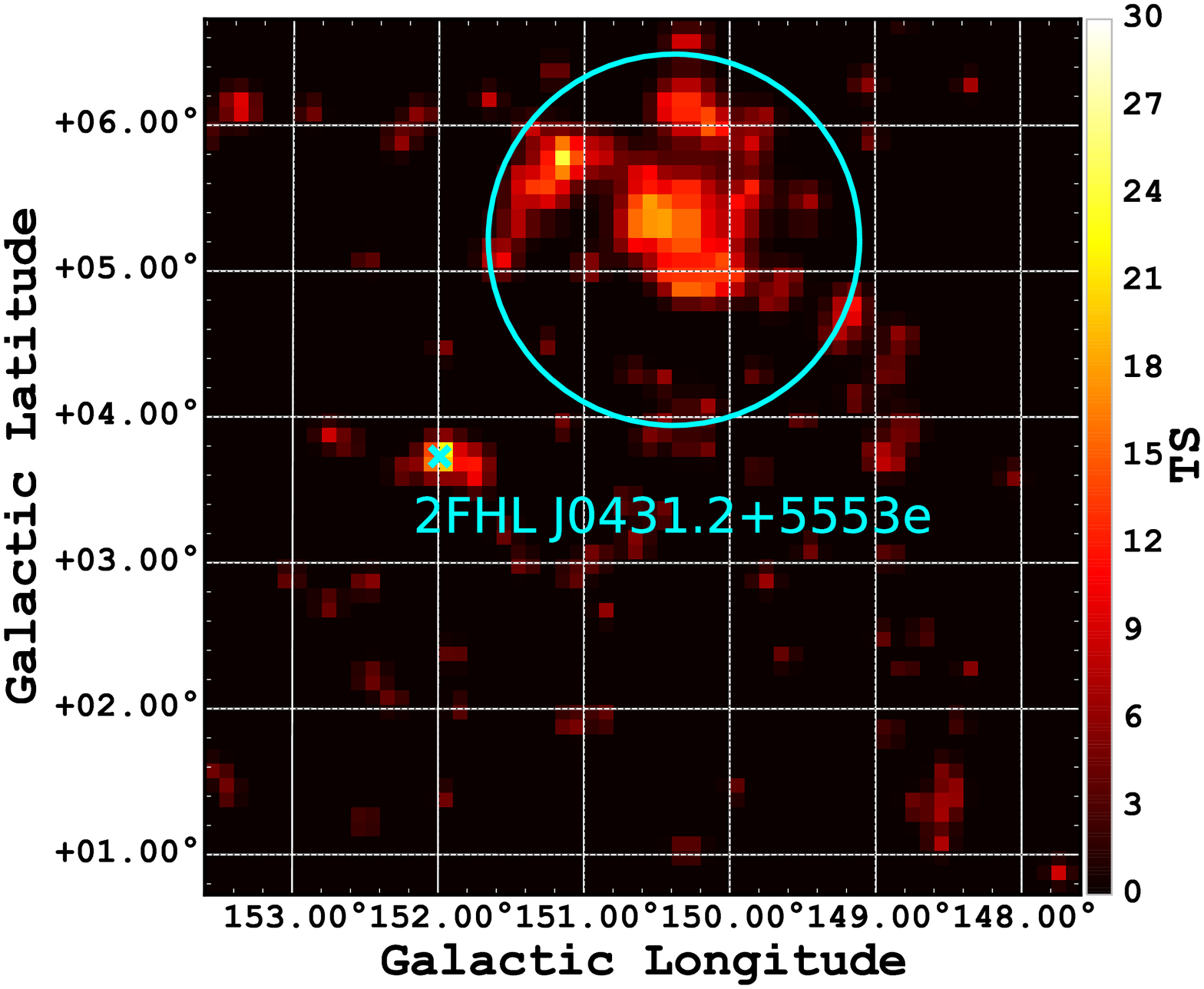}

    \caption{TS maps for the five new extended sources described in $\S$ \ref{sec:ESresults}.  Only the Galactic diffuse and isotropic emission are included in the model to highlight the location of emission not associated with the diffuse background. Circles indicate the extents of the fit disks. {The x marker in the bottom panel (2FHL~J0431.2+5553e) shows the location of a point source in the ROI.}
    \label{fig:6ES}}
    
\end{centering}
\end{figure*}

A brief description of the five new 2FHL extended sources is given below with residual TS maps for the region surrounding each source shown in Figure \ref{fig:6ES}. Detailed analyses of these new extended sources will be reported in separate papers.

{\bfseries 2FHL~J1443.2$-$6221e} overlaps with the young, radio-detected SNR RCW 86 (G315.4−2.3). RCW 86 is a 42$'$ diameter SNR that lies at a distance of 2.3-2.8 kpc and is likely associated with the first recorded supernova, SN 185 AD \citep{Rosado96,Sollerman03}. With more than 40 months of data and using the {\tt P7SOURCE} dataset, the LAT did not significantly detect the SNR, but upper limits on detection at GeV energies combined with detection of significant extension in the TeV \citep{Aharonian09} were sufficient to strongly favor a leptonic origin for the emission \citep{Lemoine-Goumard12}.

An updated LAT analysis of RCW~86 using 76 months of data, as well as the Pass 8 event-level analysis, resulted in detection of the SNR by the LAT as well as significant extension measurement \citep{Hewitt15}. In this paper, we report the results derived for 2FHL~J1443.2$-$6221e from the {\tt pointlike} analysis described in $\S$ \ref{sec:3FGL_ES}.

{\bfseries 2FHL~J1419.2$-$6048e} is a newly detected extended sources with size
${\rm \sigma_{disk} =  0.36 ^{\circ} \pm}$ $0.03 ^{\circ}$, that overlaps two nearby PWN/PSR complexes in the Kookaburra region. In the southwest of Kookaburra, HESS~J1418$-$609 \citep{AharonianKook06} is coincident with both the extended non-thermal X-ray ``Rabbit" PWN \citep[G313.3+0.1,][]{Roberts99}, and the $\gamma$-ray detected pulsar PSR~J1418$-$6058 \citep{AbdoBlindPSR09}. The northeast region, called ``K3", contains HESS~J1420$-$607, coincident with PWN~G313.5+0.3 and PSR J1420$-$6048. \cite{Acero13} detected, with \lat, emission from both HESS~J1418$-$609 (with a soft spectral index, pulsar-like spectrum) and HESS~J1420$-$607 (with a hard power-law index) above 10 GeV, but only HESS J1420$-$607 was significantly detected above 30 GeV. Neither showed significant extension. Our result for the fitted power-law spectral index of 2FHL~J1419.2$-$6048e is in agreement with the previous GeV and TeV results, yet our measured radius is considerably larger than the TeV extension. To compare the extensions of the uniform disk model used for 2FHL~J1419.2$-$6048e in this paper to the Gaussian model of \cite{AharonianKook06}, we defined the radius which contains 68\% of the source's intensity as r$_{68}$, with ${\rm r_{68,Gaussian} = 1.51\sigma}$, and ${\rm r_{68,disk} = 0.82\sigma}$  \citep{Lande12}. We find that ${\rm r_{68} \simeq 0.30^{\circ}}$ for 2FHL~J1419.2$-$6048e, and ${\rm r_{68}} \simeq 0.09^{\circ}$ for HESS~J1420$-$607.

{\bfseries 2FHL J1355.2$-$6430e}, coincident with the VHE source HESS J1356$-$645, is detected as extended (${\rm \sigma_{disk} =  0.57^{\circ} \pm 0.02^{\circ}}$) for the first time by the LAT in this work. The  source HESS J1356$-$645 \citep{Abramowski11} is associated with the pulsar PSR J1357$-$6429, which was determined to be powering a surrounding extended radio and X-ray PWN \citep{Lemoine-Goumard11}. \cite{Acero13} detected faint emission from the nebula, and derived a 99\% c.l. Bayesian upper limit on extension (${\rm \sigma_{Gauss} < 0.39^{\circ}}$) in the absence of significant extension. The fitted spectral index for 2FHL J1355.2$-$6430e is compatible with the GeV and TeV results \citep{Acero13,Abramowski11}, however, the fitted disk extension is larger than that of the TeV detection, with ${\rm r_{68} \simeq 0.47^{\circ}}$ for 2FHL~J1355.2$-$6430e and ${\rm r_{68}}$ $\simeq 0.30^{\circ}$ for HESS~J1356$-$645.

{\bfseries 2FHL J1112.4$-$6059e} is an extended source (${\rm \sigma_{disk} =  0.53^{\circ} \pm 0.03^{\circ}}$) newly detected by the LAT that encircles two 3FGL sources, 3FGL J1111.9$-$6058 and 3FGL J1111.9$-$6038, and has another, 3FGL J1112.0$-$6135, just outside its boundary \citep{3FGL}. The extended source also partially overlaps the massive star forming region NGC 3603.

Finally, {\bfseries 2FHL J0431.2+5553e} is a large extended source (${\rm \sigma_{disk} =  1.27^{\circ} \pm 0.04^{\circ}}$), with { a hard spectrum}, that has not been previously detected at $\gamma$-ray energies. It overlaps the recently discovered radio SNR G150.3+4.5 \citep{Gao14}. G150.3+4.5 is a ${\rm 2.5^{\circ}}\times {\rm 3^{\circ}}$ (Galactic coordinates)
elliptical shell type SNR that has a steep radio synchrotron spectrum ($\alpha = -0.6$), indicative of radio SNRs.

\begin{deluxetable}{lccclccc}
\setlength{\tabcolsep}{0.04in}
\tablewidth{0pt}
\tabletypesize{\scriptsize}
\tablecaption{2FHL extended sources previously detected by the {\it Fermi}-LAT \label{tab:extended}}
\tablehead{
\colhead{2FHL Name} & 
\colhead{$l$ [deg]} & 
\colhead{$b$ [deg]} &
\colhead{TS} &
\colhead{Association} &
\colhead{Class} &
\colhead{Spatial model} &
\colhead{Extension [deg]}
}
\startdata
 J0526.6$-$6825e      &    278.843 &    -32.850 & 49.80  & LMC                & gal    & 2D Gaussian & 1.87 \\
 J0617.2+2234e        &    189.048 &      3.033 & 398.64 & IC~443             & snr    & 2D Gaussian & 0.27 \\
 J0822.6$-$4250e      &    260.317 &	 -3.277 &  63.87 & Puppis A	      & snr    & Disk	     & 0.37 \\
 J0833.1$-$4511e      &    263.333 &     -3.104 & 49.70  & Vela~X             & pwn    & Disk        & 0.91 \\
 J0852.8$-$4631e      &    266.491 &     -1.233 & 437.21 & Vela~Jr            & snr    & Disk        & 1.12 \\
 J1303.4$-$6312e      &    304.235 &     -0.358 & 56.06  & HESS~J1303$-$631   & pwn    & 2D Gaussian & 0.24 \\
 J1514.0$-$5915e      &    320.269 &     -1.276 & 165.51 & MSH~15$-$52        & pwn    & Disk        & 0.25 \\
 J1615.3$-$5146e      &    331.659 &     -0.659 & 128.15 & HESS~J1614$-$518   & spp    & Disk        & 0.42 \\
 J1616.2$-$5054e      &    332.365 &     -0.131 & 87.18  & HESS~J1616$-$508   & pwn    & Disk        & 0.32 \\
 J1633.5$-$4746e      &    336.517 &      0.121 & 114.17 & HESS~J1632$-$478   & pwn    & Disk        & 0.35 \\
 J1713.5$-$3945e      &    347.336 &     -0.473 & 60.98  & RX~J1713.7$-$3946  & snr    & Map         & 0.56 \\
 J1801.3$-$2326e      &      6.527 &     -0.251 & 50.20  & W~28               & snr    & Disk        & 0.39 \\
 J1805.6$-$2136e      &      8.606 &     -0.211 & 160.43 & W~30               & snr    & Disk        & 0.37 \\
 J1824.5$-$1350e      &     17.569 &     -0.452 & 266.09 & HESS~J1825$-$137   & pwn    & 2D Gaussian & 0.75 \\
 J1834.9$-$0848e      &     23.216 &     -0.373 &  67.30 & W~41               & snr    & 2D Gaussian & 0.23 \\
 J1836.5$-$0655e      &     25.081 &      0.136 & 62.72  & HESS~J1837$-$069   & pwn    & Disk        & 0.33 \\
 J1840.9$-$0532e      &     26.796 &     -0.198 & 163.15 & HESS~J1841$-$055   & pwn    & Elliptical  & 0.62, 0.38, 39.0 \\
 J1923.2+1408e        &     49.112 &     -0.466 & 44.60  & W~51C              & snr    & Elliptical  & 0.38, 0.26, 90.0 \\
 J2021.0+4031e        &     78.241 &      2.197 & 115.97 & Gamma Cygni        & snr    & Disk        & 0.63 \\
 J2028.6+4110e        &     79.601 &      1.396 & 28.09  & Cygnus Cocoon      & sfr    & 2D Gaussian & 3.0 \\
\enddata
\tablecomments{~List of the 20 extended sources in the 2FHL that were previously detected as extended by the {\it Fermi}-LAT. All these sources are in  3FGL except W41, which is studied by \citet{W41}. The Galactic coordinates $l$ and $b$ are given in degrees. The extension of the disk templates is given by the radius, the extension of the 2D Gaussian templates is given by the $1\sigma$ radius, and the elliptical templates are given by the semi-major axis, semi-minor axis, and position angle (East of North).
}
\end{deluxetable}

\begin{deluxetable}{lcccccccclccc}
\setlength{\tabcolsep}{0.04in}
\tablewidth{0pt}
\tabletypesize{\scriptsize}
\tablecaption{New 2FHL extended sources \label{tab:new_extended}}
\tablehead{
\colhead{2FHL Name} & 
\colhead{$l$ [deg]} & 
\colhead{$b$ [deg]} &
\colhead{TS} & 
\colhead{TS$_{ext}$} &
\colhead{TS$_{2pts}$} &
\colhead{$F_{50}$} & 
\colhead{$\Delta F_{50}$} &
\colhead{$\Gamma$} & 
\colhead{$\Delta \Gamma$} &
\colhead{Association} &
\colhead{Class} &
\colhead{Radius [deg]} 
}
\startdata
 J0431.2+5553e        &    150.384 &      5.216 &  87.9 & 83.4  & 26.2    &  11.70 &       2.11 &    1.66 &         0.20 & G~150.3+4.5     & snr     & 1.27 \\
 J1112.4$-$6059e      &    291.222 &     -0.388 &  80.9 & 68.3   & 22.5    &  12.80 &       2.36 &    2.15 &         0.28 & PSR~J1112$-$6103  & pwn     & 0.53 \\
 J1355.2$-$6430e      &    309.730 &     -2.484 &  82.3 & 31.8   & 12.9     &  9.59  &       1.95 &    1.56 &         0.22 & PSR~J1357$-$6429  & pwn     & 0.57 \\
 J1419.2$-$6048e      &    313.432 &      0.260 & 109.3 & 49.1   & 15.6    &  17.60 &       2.80 &    1.87 &         0.19 & PSR~J1420$-$6048  & pwn     & 0.36 \\
 J1443.2$-$6221e      &    315.505 &     -2.239 &  75.6 & 29.9   & 19.2   &  7.23  &       1.70 &    2.07 &         0.30 & SNR~G315.4$-$2.3  & snr     & 0.27 \\
\enddata
\tablecomments{~List of the 5 new extended sources in the 2FHL. All these sources are characterized by an uniform disk template whose radius is given in the last column.}
\end{deluxetable}

%%%%%%%%%%%%%%%%%%%%%%%%%%%%%%%%%%%%%%%%%%%%%%%%%%%%%%%%%%%%%%%%
%
%         Extragalactic Science
%
%%%%%%%%%%%%%%%%%%%%%%%%%%%%%%%%%%%%%%%%%%%%%%%%%%%%%%%%%%%%%%%%
\subsection{The 2FHL Extragalactic Sky}
\subsubsection{General and spectral properties}
Most of the sources detected in 2FHL are extragalactic. 83\,\% (299)  are either located at \blot or associated with an extragalactic source. We refer to this set of sources as the extragalactic sample.

\begin{figure}[!ht]
\begin{centering}
\includegraphics[width=\columnwidth]{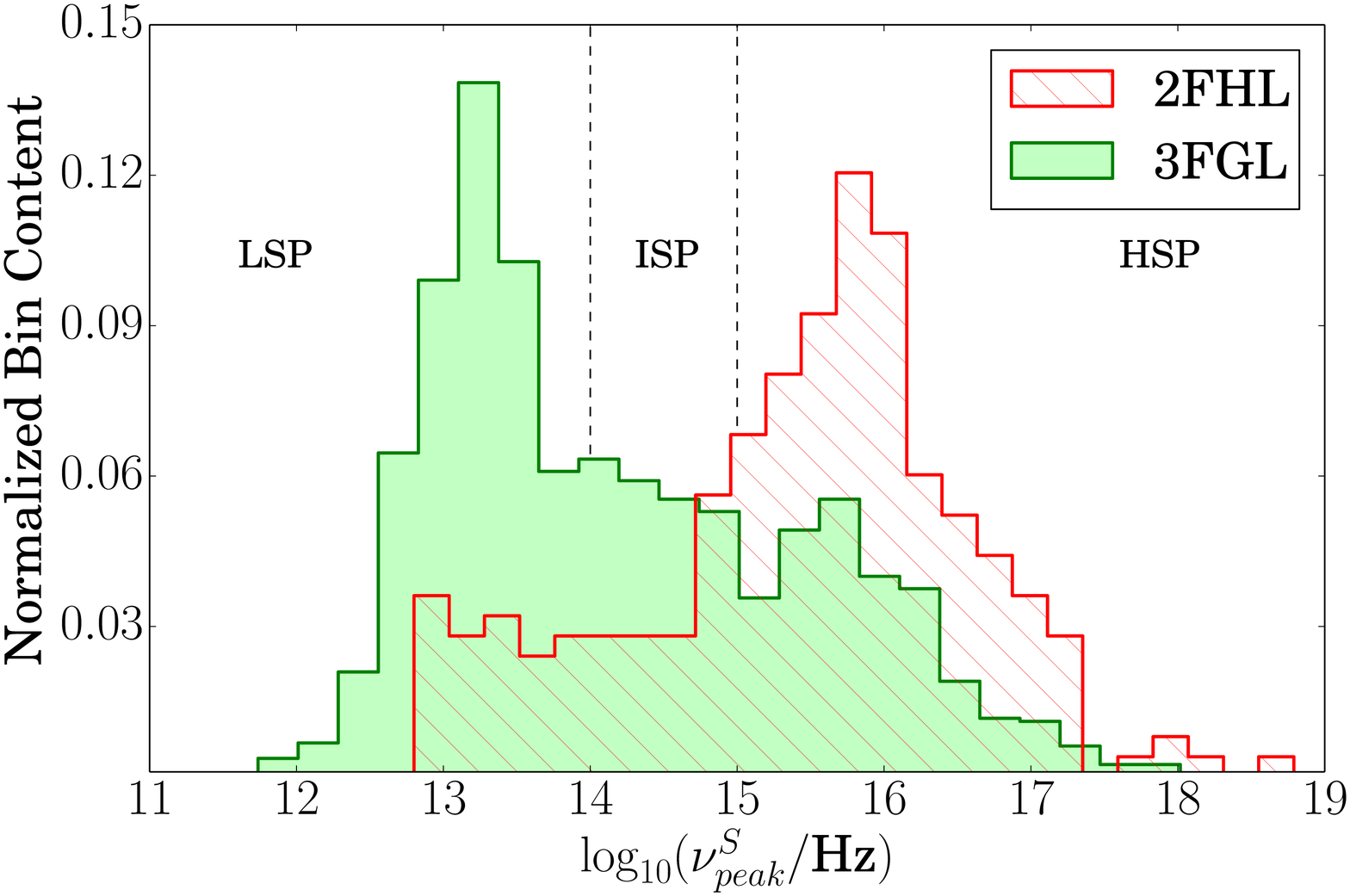} 
    \caption{Normalized distribution of the frequency of the synchrotron peak for the blazars detected in the 3FGL and those detected in the 2FHL. For the 3FGL
sources the peak frequencies were adopted from 3LAC \citep{3LAC}.
{ LSP, ISP and HSP blazars are those with
$\log_{10}(\nu_{peak}^{S}/{\rm Hz})<14$, 
$14<\log_{10}(\nu_{peak}^{S}/{\rm Hz})<15$, and
$\log_{10}(\nu_{peak}^{S}/{\rm Hz})>15$ respectively.
}
    \label{fig:index_vs_peak}}
\end{centering}
\end{figure}

BL Lacs represent the most numerous source class (54\,\% of the full 2FHL catalog and 65\% of the extragalactic sample), while there are a few detected FSRQs (10 sources, 3\,\% of the full 2FHL catalog). Such a low number of FSRQs is expected due to their soft spectra at $>$50\,GeV. 
Most of the detected BL Lacs belong to the high-frequency synchrotron peak (HSP) class,
{ rather than to the  low-frequency synchrotron peak (LSP), or 
intermediate-frequency synchrotron peak (ISP) class.}
This is shown in Figure~\ref{fig:index_vs_peak}, which reports the distribution of synchrotron peak frequencies of blazars detected in the 2FHL and in the 3FGL. It is clear that the two catalogs sample different parts of the blazar population, with the 3FGL including mostly LSPs and ISPs and the 2FHL including mostly HSPs.

There are 198 2FHL sources in the extragalactic sample that are detected both in the 3FGL and 1FHL catalogs. { There are 33 other sources that are  neither in the 3FGL nor in the 1FHL catalog (and not in 1FGL/2FGL either).}
  In general, 2FHL sources
not detected in 3FGL and 1FHL are harder (median $\Gamma \sim$2.8 versus 3.2),
fainter (median $F_{50}\sim 1.3\times 10^{-11}$\,ph cm$^{-2}$ s$^{-1}$ versus $2.3\times 10^{-11}$\,ph cm$^{-2}$ s$^{-1}$) and are detected at lower
significance ({median} TS$\sim 30$ versus $53$) than those detected in the 
aforementioned catalogs.

One of the 33 exclusive 2FHL sources is 2FHL J1944.1+2117 ($\Gamma=2.73\pm 0.66$) that is associated with HESS J1943+213 \citep{abramowski11b}. This source is potentially an extreme high-frequency peaked BL Lac, classified as { bcu II}, located in the Galactic plane with $\log_{10}(\nu_{peak}^{S}/{\rm Hz})=18.79$ \citep{3LAC}. We also detect N~157B, 2FHL J0537.4$-$6908, with a very hard spectral index of $\Gamma=1.15\pm 0.37$. This is an extragalactic PWN in the LMC detected by H.E.S.S. up to 18~TeV \citep{abramowski12,abramowski15}. The other 31 sources (out of the 33 2FHL exclusive sources) are not detected by Cherenkov telescopes yet. Of the 31 sources, 16 are classified as some type of blazar whereas 14 are unclassified. One source  is associated with a galaxy cluster, 2FHL J0318.0$-$4414 detected with a soft spectrum ($\Gamma=4.0\pm 1.7$). The emission likely originates from  PKS~0316$-$444, which is a bright radio source at the center of  Abell~3112 \citep[$z=0.075$,][]{takizawa03}. Emission from the region of Abell~3112 was already marginally found by \citet{ackermann14}, yet it is robustly detected in the 2FHL.

We find that the number of FSRQs is strongly reduced from 1FHL (71 sources) to the 2FHL (10 sources). This is mainly due to the softer indices and spectral curvature that characterize the spectrum of this blazar population at these higher energies. Another indication of this effect is that only 2 sources at $z>1$ (out of a total of 7) are FSRQs, { while the rest are BL Lacs}. Other noteworthy examples of detected extragalactic sources are the nearby radio galaxies IC 310 ($z=0.019$), NGC~1275 ($z=0.0175$), PKS~0625-35 ($z=0.05494$), 3C~264 ($z=0.021718$), M~87 ($z=0.004283$) and Centaurus A ($z=0.0018$).

\begin{figure*}[!ht]
 \begin{centering}
\includegraphics[width=8cm]{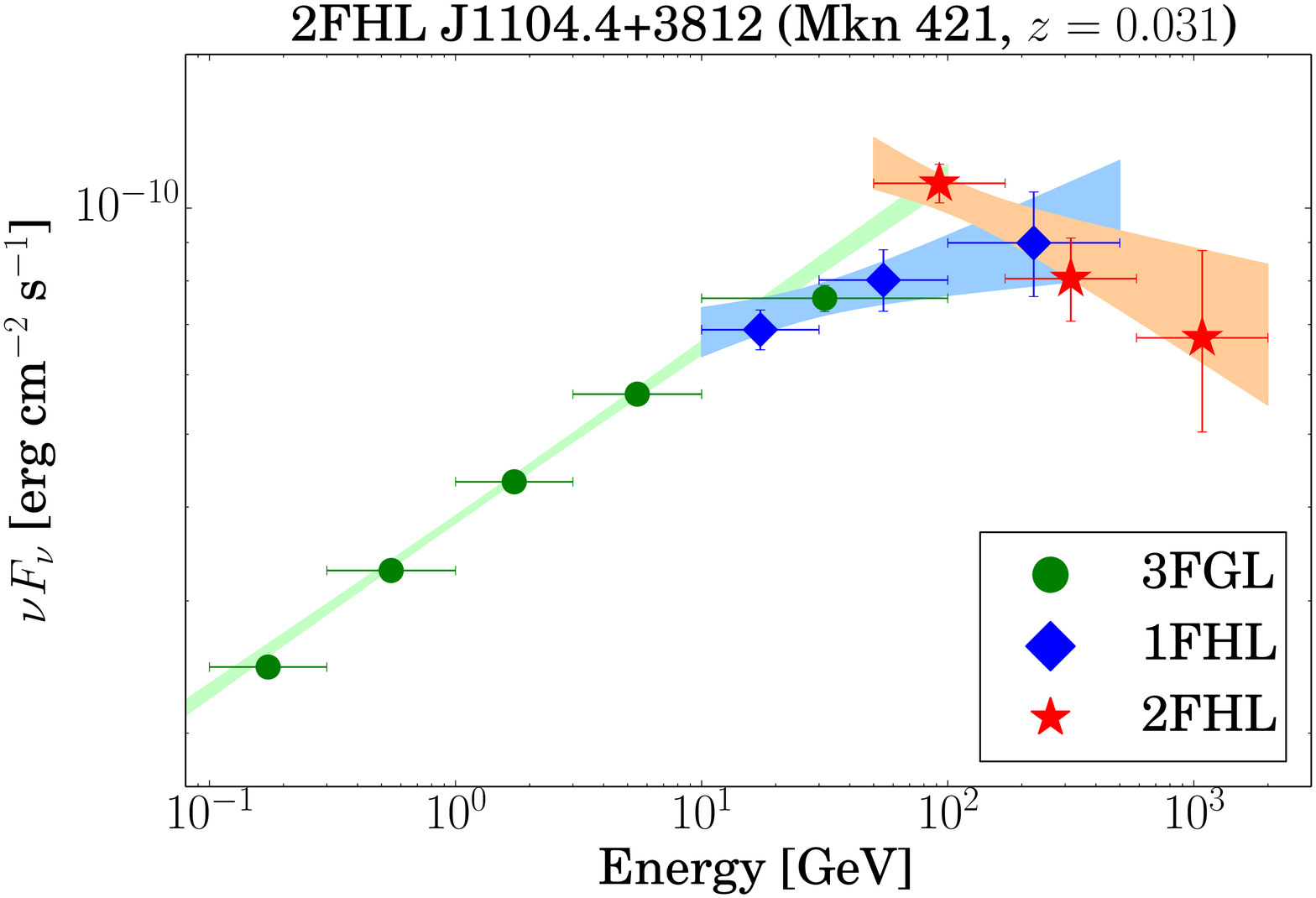}
\includegraphics[width=8cm]{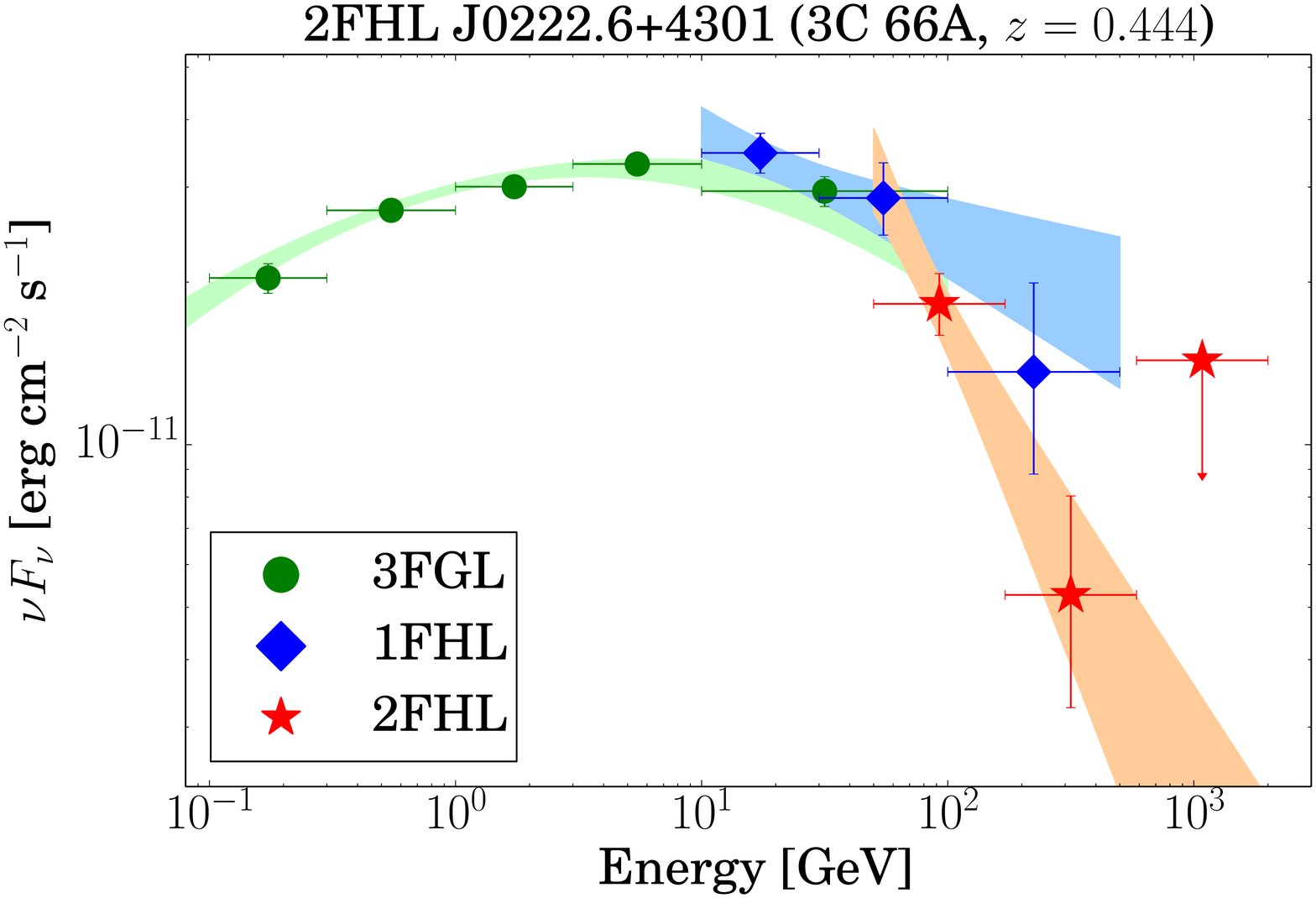}
\caption{The spectral energy distributions of Mrk~421 ({\it left panel}) and 3C~66A ({\it right panel}) over four decades in energy. The higher energy peak is well characterized by combining data from the 3FGL (green diamonds), 1FHL (blue circles), and 2FHL (red stars).
\label{fig:seds}}
\end{centering}
\end{figure*}

\begin{figure}[!ht]
\begin{centering}
    \includegraphics[width=\columnwidth]{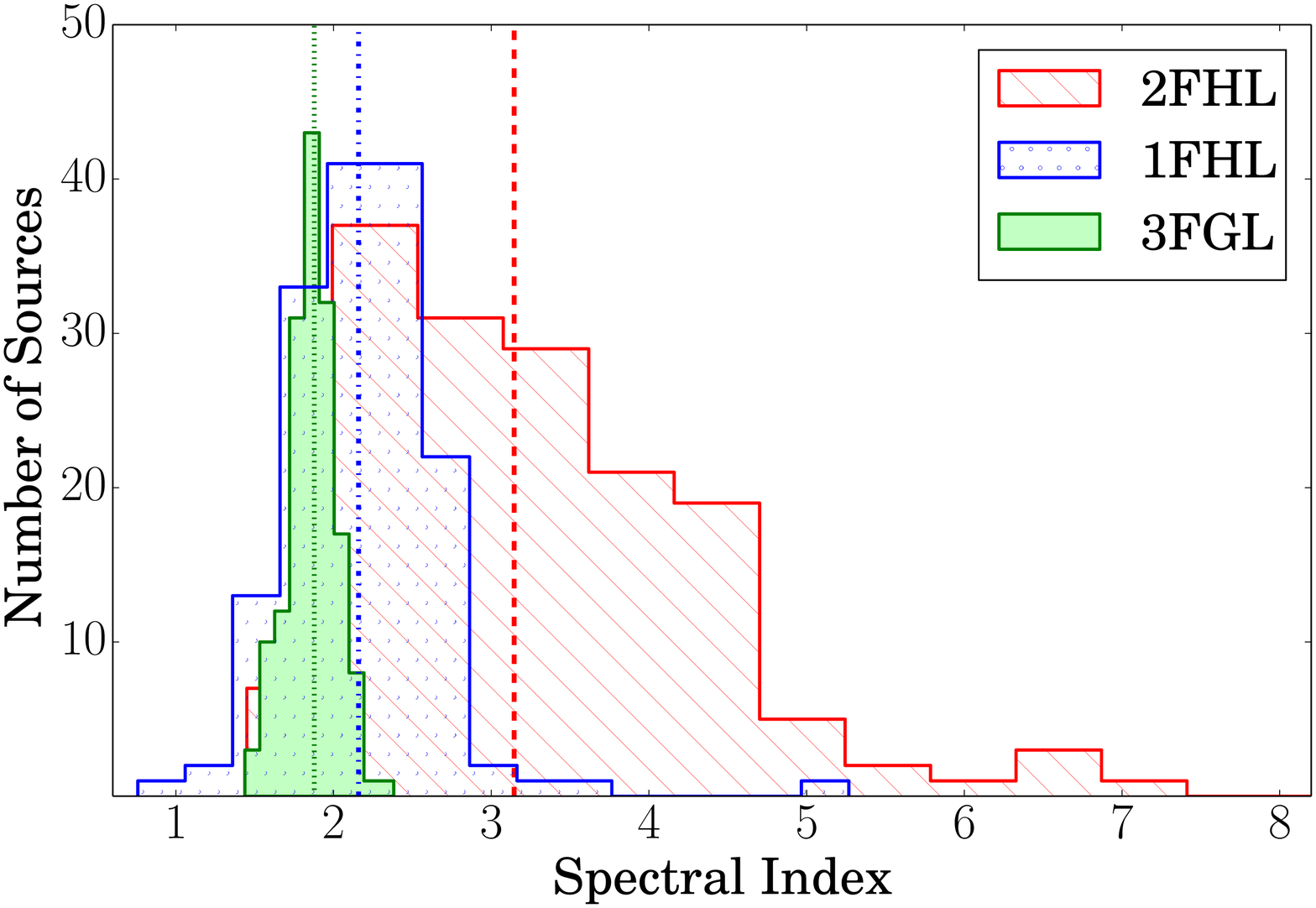} 
    \caption{The distribution of spectral indices for a subsample of 158 BL Lacs that are in common among the 2FHL (backslash orange), 3FGL (slash green), and 1FHL (purple). The medians of the distributions are shown with vertical lines. The higher the energy band, the larger the index; therefore sources get softer with increasing energy. The scatter of the distribution is also larger with increasing energy, partly because of the lower statistics.
    \label{fig:drop}}
\end{centering}
\end{figure}

Figure~\ref{fig:seds} shows the SEDs of two notable sources, Mrk 421 and 3C 66A, highlighting how the LAT now resolves the descending part of the high-energy peak. Indeed, 2FHL spectra of extragalactic sources are generally softer than the corresponding 3FGL and 1FHL spectra. This is evident in Figure~\ref{fig:drop}, which compares the spectral indices for a subsample of 158 BL Lacs in common between the 2FHL, 3FGL and 1FHL catalogs. There are two interesting facts illustrated by  Figure~\ref{fig:drop}. First, the median of the distribution shifts towards larger (softer) indices since {2FHL} samples the drop of the SED. Second, the scatter of the distribution becomes larger with increasing energies. 
Some of the scatter arises from statistical effects since the number of detected photons at $>$50\,GeV is generally lower than in the 3FGL and 1FHL.

\subsubsection{$\gamma$-ray attenuation}
\begin{figure}[!ht]
\begin{centering}
\includegraphics[width=\columnwidth]{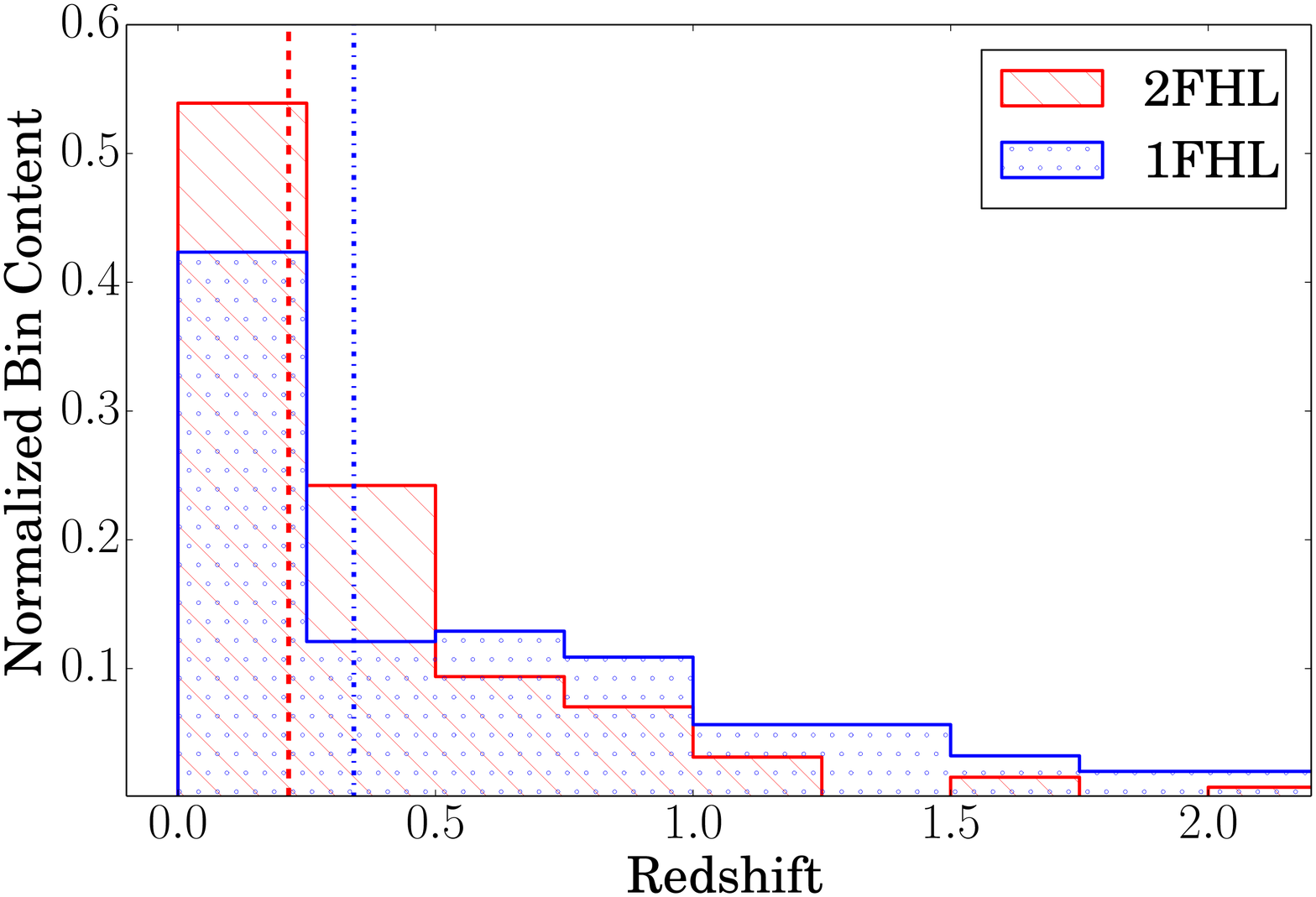} 
    \caption{Normalized redshift distribution of the sources with known redshift in 2FHL (orange backslash) and 1FHL (blue dotted). The medians of the distributions are plotted with a dashed and dash-dotted vertical line, respectively. 
    \label{fig:hist_z}}
\end{centering}
\end{figure}

\begin{figure}[!ht]
\begin{centering}
\includegraphics[width=\columnwidth]{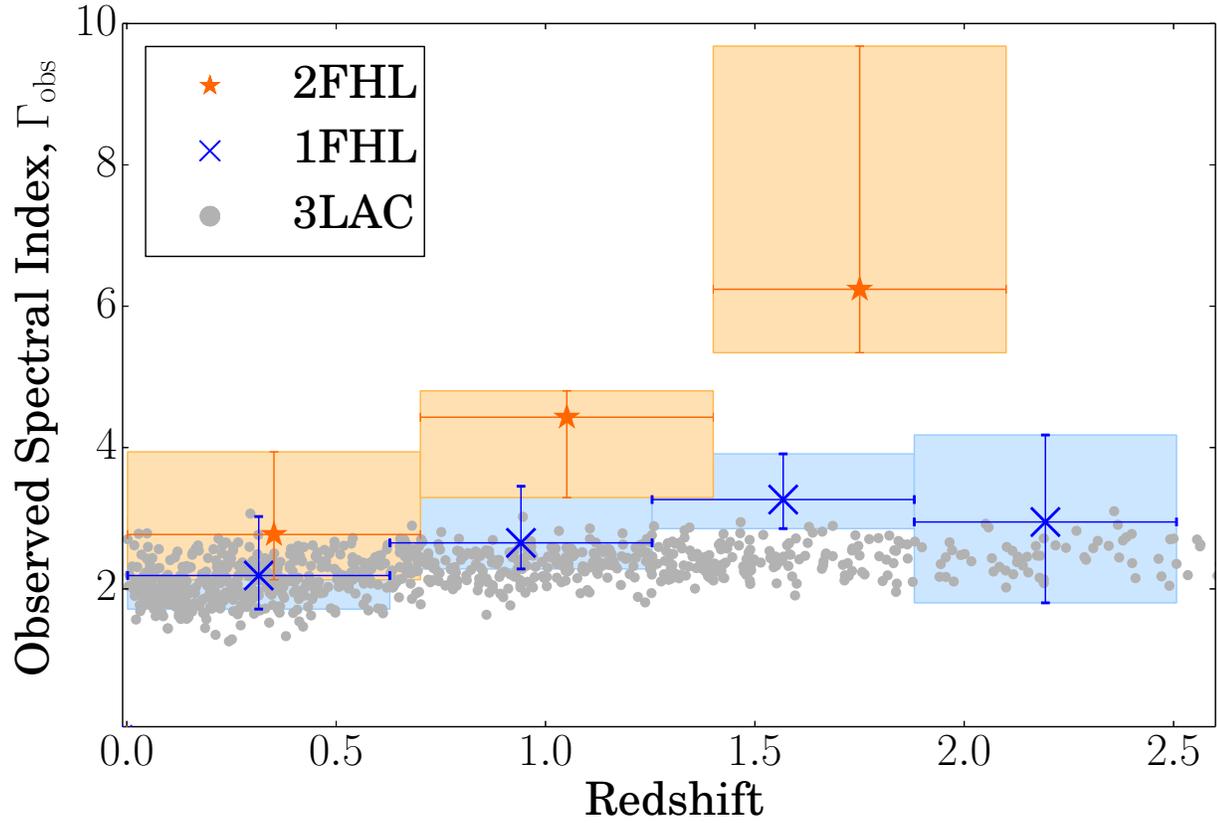}
    \caption{Observed spectral index versus redshift of the 3LAC sources (energy range, 0.1\,GeV--100\,GeV), the median spectral index of the 1FHL sources (10~GeV--500~GeV) in 4 redshift bins, and the median spectral index of the 2FHL sources (50~GeV--2~TeV) in 3 redshift bins. The uncertainties are calculated as the 68\,\% containment around the median.
There is a dependence of the spectral index on redshift at the energies where the EBL attenuation is significant.
    \label{fig:index_vs_z}}
\end{centering}
\end{figure}

Spectroscopic redshift measurements exists in the literature \citep[e.g.][]{shaw12,shaw13,masetti13} for 43\,\% (128) of the 299 extragalactic sources. The comparison of the 2FHL and 1FHL redshift distributions (reported in Figure~\ref{fig:hist_z}) shows that on average 2FHL sources lie at lower redshifts than the 1FHL ones. This might be because FSRQs and LSP blazars, which tend to be located at higher redshifts, have  soft spectra and are faint in the 2FHL band. High redshift sources appear even fainter because of the extragalactic background light (EBL) attenuation. The EBL contains all the photons, from the ultraviolet to the infrared, that have been emitted by star formation processes and supermassive black hole accretion throughout the history of the Universe  \citep[\eg][]{dwek13}. The spectrum of extragalactic $\gamma$-ray sources, in the whole energy range of our analysis, is modified by pair-production interaction of the $\gamma$-ray photons with the EBL \citep[\eg][]{ebl12}. This interaction produces an attenuation { of} the observed fluxes that is energy and redshift dependent (the larger the $\gamma$-ray energy and/or the redshift, the larger the attenuation). We note that despite the  high energies probed by this work, seven sources have $z>1$, where the EBL attenuation is not negligible. These sources are associated with PKS~0454$-$234 ($z=1.003$), PKS~0426$-$380 \citep[$z=1.105$,][]{tanaka13}, OJ~014 ($z=1.148$), TXS~0628$-$240 ($z=1.238$), PKS~B1424$-$418 ($z=1.55$), B2~2114+33 ($z=1.6$) and MG4~J000800+4712 ($z=2.1$). We specially note that three sources have redshifts $z>1.5$. The effect of the EBL attenuation is clearly seen in Figure~\ref{fig:index_vs_z}. This figure shows the significant dependence of the spectral index on the redshift at the 2FHL energies whereas that dependence is reduced at lower energies (\ie for 3FGL and 1FHL) where the EBL interaction is less relevant.

\begin{figure}[!ht]
\begin{centering}
    \includegraphics[width=\columnwidth]{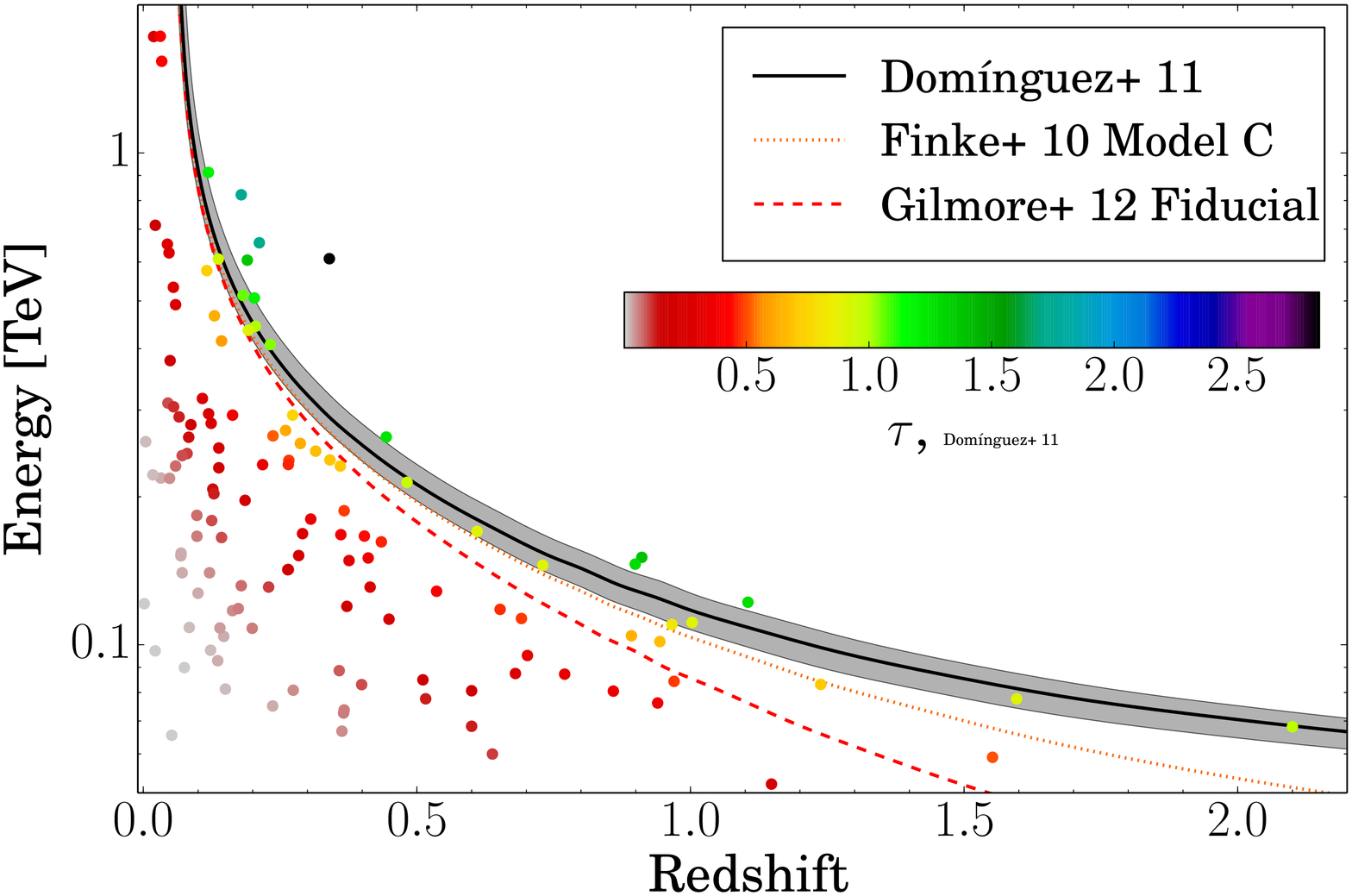}
    \caption{The highest photon energy versus source redshift. The symbols are color coded by the optical depth, $\tau$, estimated from the EBL model by \citet{dominguez11a}. Different estimates of the cosmic $\gamma$-ray horizon are plotted as well, which are derived from the EBL models by \citet[][dotted orange line]{finke10}, \citet[][solid black line, with its uncertainties as a shaded band]{dominguez11a} and \citet[][dashed red line]{gilmore12}. We note that several photons are from near and beyond the horizon.
    \label{fig:energy_vs_z}}
\end{centering}
\end{figure}

The EBL attenuation can be parameterized by an optical depth $\tau$ that can be derived from empirical EBL models \citep[\eg][]{franceschini08,kneiske10,finke10,dominguez11a,gilmore12,stecker12,helgason12,khaire15}. In fact, the EBL sets a distance limit from where $\gamma$-ray photons of a given energy are expected to reach us, the cosmic $\gamma$-ray horizon \citep[CGRH, \eg][]{lat_ebl10,dominguez13a}. Formally, the CGRH may be defined as the energy at which $\tau=1$ as a function of redshift. Figure~\ref{fig:energy_vs_z} shows the HEPs in 2FHL versus the redshift of emission. As we can see from Figure~\ref{fig:energy_vs_z}, a few photons are from near and beyond the CGRH.
These photons from significantly attenuated sources provide information about the EBL, which in turn carries fundamental information about galaxy evolution and cosmology \citep[\eg][]{hauser01,dominguez13b,biteau15}. In fact, the most complete EBL information is typically given by the sources detected up to the highest energies, which are usually the brightest sources with the hardest spectra. The HEPs with the largest optical depths are associated with RBS~0413 ($z=0.19$, $\tau=1.38$), PKS~0823$-$223 ($z=0.911$, $\tau=1.39$), RX~J0648.7+1516 ($z=0.179$, $\tau=1.74$), 1ES~0347$-$121 ($z=0.188$, $\tau=2.43$), 1ES~0502+675 ($z=0.34$, $\tau=2.84$), where $\tau$ is given by the observationally-based EBL model by \citet{dominguez11a}. These photons are especially interesting for testing alternative photon-propagation scenarios such as secondary cascades \citep{essey10} and axion-like particles \citep[\eg][]{deangelis07,sanchez-conde09,dominguez11b,horns12}. { However, we caution the reader that Figure~\ref{fig:energy_vs_z} cannot be readily interpreted in terms of constraints on the EBL optical depth because brighter and harder sources stand a larger chance to have photons detected from beyond the horizon.}

%%%%%%%%%%%%%%%%%%%%%%%%%%%%%%%%%%%%%%%%%%%%%%%%%%%%%%%%%%%%%%%%
%
%         Variability
%
%%%%%%%%%%%%%%%%%%%%%%%%%%%%%%%%%%%%%%%%%%%%%%%%%%%%%%%%%%%%%%%%
\subsection{Variability of 2FHL Sources}
\label{sec:var}

We have studied the variability of 2FHL sources using the same
Bayesian Block algorithm adopted in 1FHL \citep{1FHL}.
This algorithm, designed to detect and characterize variability in
time series \citep{scargle98,scargle13}, is well suited in regimes of low-count statistics.
The algorithm partitions a given source light curve into constant
segments (blocks), each characterized by a flux and duration.
The location of the transition between blocks is determined
by optimizing a fitness function \citep[using the algorithm of ][]{jackson05} 
for the partitions. As for the analysis of 1FHL data, the fitness
function employed here is the logarithm of the maximum likelihood
for each individual block under the hypothesis of a constant local
flux \citep[see][]{scargle13}. A 1\,\% false positive 
threshold was selected for all sources. For 360 sources, we expect 3-4 false detections.
Aperture photometry was performed for each source using an ROI of 0.5$^{\circ}$
radius centered on the maximum likelihood source coordinates. The light curves were divided into 50 equal time bins spanning the 80-month interval. For sources with neighboring 2FHL sources closer than 1$^{\circ}$, the radius of the ROI was decreased to the greater value of (half the angular separation) or 0.25$^{\circ}$ . No background subtraction was done for the aperture photometry and Bayesian Block analyses. This is the same procedure that was used in 1FHL. Five pairs of sources are closer than 0.5$^{\circ}$, all of which are located in the Galactic plane.

\begin{figure*}[!ht]
\begin{centering}
    \includegraphics[angle=0,width=\textwidth]{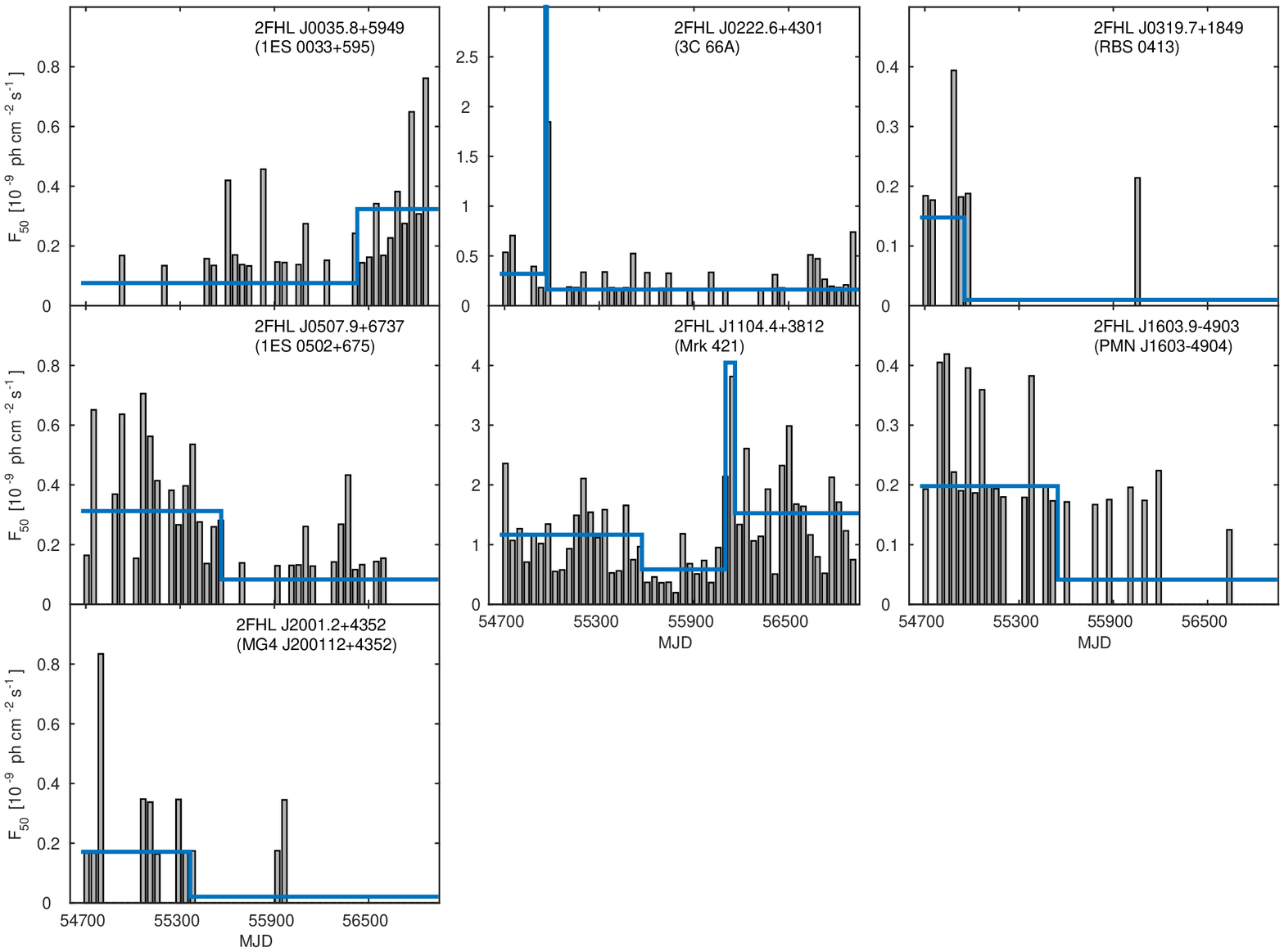} 
    \caption{Light curves for the seven variable sources detected in the 2FHL catalog. The histograms correspond to the aperture photometry analysis, and the solid lines correspond to the Bayesian Block analysis using a 1\% false positive threshold. The panels are labeled with the 2FHL names and the names of the corresponding associated sources (in parentheses). For 3C~66A, the peak flux in the Bayesian Block light curve is $5.4\times 10^{-9}$~ph~cm$^{-2}$~s$^{-1}$; the ordinate is manually truncated to $3\times 10^{-9}$~ph~cm$^{-2}$~s$^{-1}$ to improve readability. 
    \label{fig:bb_lightcurves}}
\end{centering}
\end{figure*}

Only 7 sources are detected as being variable at more than 99\,\% C.L (see 
Table~\ref{tab:var}) in contrast { to the detection of 43 variable 
sources  in 1FHL.} Most of the 2FHL variable sources
are associated with BL Lacs and all but one \citep[PMN~J1603$-$4904, associated
possibly with a young radio galaxy,][]{mueller15} have
already been detected by IACTs. Most sources
are detected with only 2 blocks, meaning that they were brighter
during part of the 80\,months spanned by the 2FHL analysis.
MG4~J200112+4352, 1ES~0502+675,  RBS~0413, and PMN~J1603$-$4904 
were all in high states at the beginning of the {\it Fermi} mission,
while 1ES 0033+595 became brighter after 2013. 3C~66A had a clear outburst
at $>$50\,GeV, coincident with the bright flare detected by VERITAS
and {\it Fermi} in September 2008 \citep{3c66a}.
The most frequently variable source (detected with 4 blocks) is Mrk~421,
which was not detected as variable in 1FHL. The Bayesian Block analysis
clearly detects the outburst of the source  in August 2012 \citep{hovatta15}. 
Figure \ref{fig:bb_lightcurves} shows the light curves for these 7 sources.
Half of the sources in 2FHL are detected with less than 6 photons, which prevents us from assessing whether variability is present. Indeed, the weakest source for which we detect  variability is 2FHL~J0319.7+1849 (RBS~0413), which is detected with $\sim$6 photons.

\begin{deluxetable}{lclcc}
\setlength{\tabcolsep}{0.04in}
\tablewidth{0pt}
\tabletypesize{\small}
\tablecaption{Results of the Bayesian Block Variability Analysis \label{tab:var}}
\tablehead{
\colhead{2FHL Name} &
\colhead{Number of Blocks} &
\colhead{$\gamma$-ray Association} &
\colhead{Class} &
\colhead{TeV?}
}
\startdata

J0035.8+5949   & 2 & 1ES~0033+595     & bll & Yes \\
J0222.6+4301   & 3 & 3C~66A           & bll & Yes \\
J0319.7+1849   & 2 & RBS~0413         & bll & Yes \\
J0507.9+6737   & 2 & 1ES~0502+675     & bll & Yes \\
J1104.4+3812   & 4 & Mrk~421          & bll & Yes \\
J1603.9$-$4903 & 2 & PMN~J1603$-$4904   & rdg & No \\
J2001.2+4352   & 2 & MG4~J200112+4352 & bll & Yes \\
\enddata
\end{deluxetable}

%%%%%%%%%%%%%%%%%%%%%%%%%%%%%%%%%%%%%%%%%%%%%%%%%%%%%%%%%%%%%%%%
%
%         Detectability by TeV telescopes
%
%%%%%%%%%%%%%%%%%%%%%%%%%%%%%%%%%%%%%%%%%%%%%%%%%%%%%%%%%%%%%%%%
\subsection{Candidates for detection with Cherenkov telescopes}
\label{sec:tevcat}
\begin{figure*}[!ht]
\begin{centering}
    \includegraphics[angle=0,width=\textwidth]{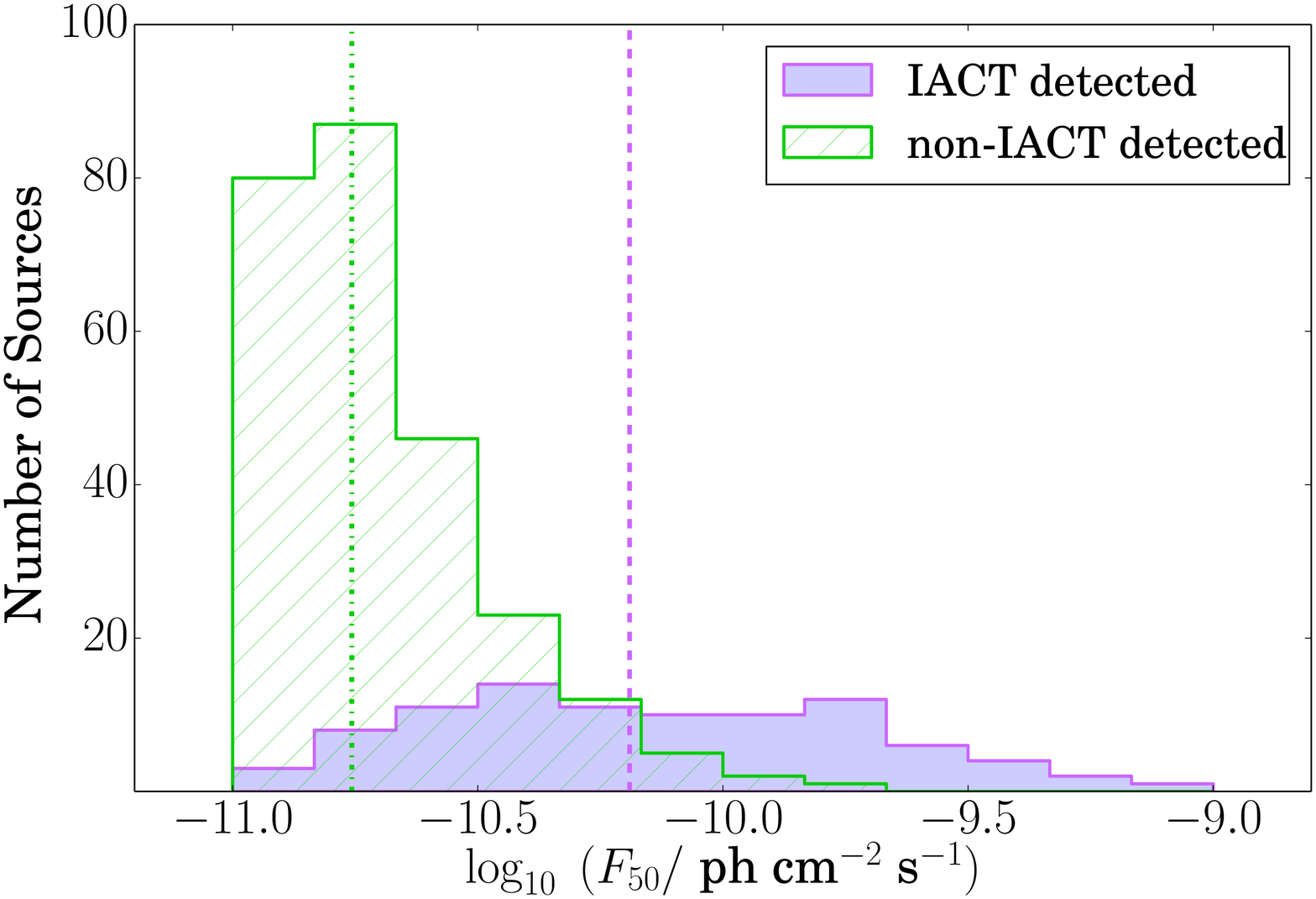} 
    \caption{Photon flux distributions of the 2FHL population that has been detected already by IACTs (purple) and that has not yet been detected (green). The medians of the distributions are shown with vertical lines.
    \label{fig:hist_TeV}}
\end{centering}
\end{figure*}

A cross correlation between the 2FHL and TeVCat catalogs, finds that 282 sources in 2FHL (i.e. $\sim$78\,\% of the 2FHL sample)  have not been detected yet by IACTs. Figure~\ref{fig:hist_TeV} shows the photon flux distributions of the population already detected by IACTs (median of $5.72\times 10^{-10}$~ph~cm$^{-2}$~s$^{-1}$) and that not detected (median of $1.55\times 10^{-10}$~ph~cm$^{-2}$~s$^{-1}$); the flux is a factor $\sim$4 greater for the population already detected. Out of  282 sources that are not detected by IACTs, 216  are located at \blot and 66  are at $|b|<10^{\circ}$. Therefore, most of the 2FHL targets for IACTs are likely extragalactic. Given the energy threshold and sensitivity of current IACTs, a large majority of sources in our catalog are promising targets for IACT detections. {However},  for extragalactic studies, it may be convenient to concentrate on HSP blazars, which typically have their higher-energy peak at energies above 100\,GeV. With a factor of 10 improvement in sensitivity and the extension to energies below 100\,GeV, all 2FHL sources should be detectable by the future Cherenkov Telescope Array \citep{acharya13}.

%%%%%%%%%%%%%%%%%%%%%%%%%%%%%%%%%%%%%%%%%%%%%%%%%%%%%%%%%%%%%%%%
%
%         Summary
%
%%%%%%%%%%%%%%%%%%%%%%%%%%%%%%%%%%%%%%%%%%%%%%%%%%%%%%%%%%%%%%%%
\section{Summary}
\label{sec:summary}

We have presented an all-sky analysis at $\geq$50\,GeV
of 80\,months of \lat data relying on the new Pass~8 event-level analysis.
Pass~8 delivers improvements in the acceptance and the PSF, reduces
background of misclassified charged particles
and extends the energy range at which \lat is sensitive.
All this allowed \lat to detect 360 sources in the 50\,GeV--2\,TeV range,
performing an unbiased census of the $>$50\,GeV sky for the first time.
This catalog of sources (dubbed 2FHL) provides a bridge between
the traditional 0.1--100\,GeV band of \lat catalogs \citep{3FGL}
and the $\gtrsim$100\,GeV band probed by IACTs from the ground.
The 2FHL catalog has the potential to improve the efficiency with
which new sources are detected at TeV energies since only
about 25\,\% of the 2FHL sources were previously detected by IACTs.

The majority ($\gtrsim$80\,\%)
of sources detected in the 2FHL catalog are likely extragalactic
because they are either located at high Galactic latitude or are associated with blazars. BL Lacs represent the largest population of sources detected by the LAT 
(54\,\%), followed by blazars of uncertain classification (16\,\%) and unassociated
sources (13\,\%). Most BL Lacs in 2FHL belong to the HSP class and display the hardest
(among blazars) $\gamma$-ray spectra and substantial emission in the 50\,GeV--2\,TeV
band. These sources are powerful probes of the EBL and this work has shown
that \lat has detected emission from many blazars at optical depths $>$1.

The 2FHL includes 103 sources in the direction of the Galactic plane ($|b|<10^{\circ}$).
While a fraction of the sources ($\sim$39\,\%) is associated with blazars, the rest
are Galactic and unassociated sources. Galactic sources generally display
much harder photon indices than blazars (median of $\sim$2 versus $\sim$3)
and copious TeV emission, both signs of efficient particle acceleration.
Most Galactic sources are associated with PWNe and SNRs, systems at the end
of the stellar evolution cycle, and are detected as spatially extended.
All the hard (spectral index $<2$) unassociated  sources within the plane
of our Galaxy are likely of Galactic origin, since very few blazars 
have spectra as hard.

A comparison with a similarly long dataset of Pass~7 photons shows that
Pass~8 allows the LAT to detect 35\,\% more sources. It is thus
clear that Pass~8 and the accumulated exposure allow the LAT to extend its reach
to higher energy and to open a new window on the sub-TeV sky. Sensitivity improves linearly with time in the photon-limited regime, thus 
further observations by the LAT in the coming years
will probe the $>$\,50\,GeV sky even more deeply, providing
important targets for current and future 
Cherenkov telescopes.

%%%%%%%%%%%%%%%%%%%%%%%%%%%%%%%%%%%%%%%%%%%%%%%%%%%%%%%%%%%%%%%%%%%%%%%%%%
%%%%%%%%%%%%%%%%%%%%%%%%%%%%%%%%%%%%%%%%%%%%%%%%%%%%%%%%%%%%%%%%%%%%%%%%%%
\acknowledgments
The \textit{Fermi} LAT Collaboration acknowledges generous ongoing support
from a number of agencies and institutes that have supported both the
development and the operation of the LAT as well as scientific data analysis.
These include the National Aeronautics and Space Administration and the
Department of Energy in the United States, the Commissariat \`a l'Energie Atomique
and the Centre National de la Recherche Scientifique / Institut National de Physique Nucl\'eaire et de Physique des Particules in France, the Agenzia 
Spaziale Italiana and the Istituto Nazionale di Fisica Nucleare in Italy, 
the Ministry of Education, Culture, Sports, Science and Technology (MEXT), 
High Energy Accelerator Research Organization (KEK) and Japan Aerospace 
Exploration Agency (JAXA) in Japan, and the K.~A.~Wallenberg Foundation, 
the Swedish Research Council and the Swedish National Space Board in Sweden.
Additional support for science analysis during the operations phase 
is gratefully acknowledged from the Istituto Nazionale di Astrofisica in 
Italy and the Centre National d'\'Etudes Spatiales in France.
This research has made use of the NASA/IPAC Extragalactic Database (NED) which is operated by the Jet Propulsion Laboratory, California Institute of Technology, under contract with the National Aeronautics and Space Administration. This research has made use of the SIMBAD database, operated at CDS, Strasbourg, France

{\it Facilities:} \facility{Fermi/LAT}

%%%%%%%%%%%%%%%%%%%%%%%%%%%%%%%%%%%%%%%%%%%%%%%%%% biblio
\bibliographystyle{apj}
\bibliography{biblio.bib}

\end{document}